\documentclass[superscriptaddress,reprint,amsmath,amssymb,aps,floatfix, longbibliography]{revtex4-1}
\usepackage{amsmath,gensymb,textcomp,bm,dcolumn,eurosym,array,tabu,multirow,nicefrac,color,subfigure,graphicx,upgreek,svg}
\usepackage[colorlinks, linkcolor=blue, citecolor=blue, urlcolor=blue, breaklinks=true]{hyperref}

\usepackage{mathpazo,times} 

\usepackage{graphicx}
\usepackage{dcolumn}
\usepackage{units}
\usepackage{float}
\usepackage{adjustbox, ulem}

\usepackage{fourier}
\DeclareRobustCommand{\bigO}{%
  \text{\usefont{OMS}{cmsy}{m}{n}O}%
}

\usepackage{color}
\usepackage[T1]{fontenc}
\definecolor{txtblue}{RGB}{0,72,186}
\definecolor{txtgreen}{RGB}{107,142,35}
\definecolor{txtred}{RGB}{187,102,55}

\begin{document}
\title{
Scalable authentication and optimal flooding in a quantum network
}

\author{Naomi R. Solomons}
\thanks{These authors contributed equally to this work}
\affiliation{Quantum Engineering Technology Labs \& Quantum Engineering Centre for Doctoral Training, Centre for Nanoscience and Quantum Information, University of Bristol, Bristol, United Kingdom}

\author{Alasdair I. Fletcher}
\thanks{These authors contributed equally to this work}
\affiliation{Department of Computer Science, University of York, York, YO10 5GH United Kingdom}

\author{Djeylan Aktas}
\thanks{Currently works at: \textit{RCQI, Institute of Physics, Slovak Academy of Sciences, Dúbravská Cesta 9, 84511 Bratislava, Slovakia}}
\affiliation{Quantum Engineering Technology Labs, H. H. Wills Physics Laboratory \& Department of Electrical and Electronic Engineering, University of Bristol, Merchant Venturers Building, Woodland Road, Bristol BS8 1UB, United Kingdom}
\author{Natarajan Venkatachalam}
\affiliation{Quantum Engineering Technology Labs, H. H. Wills Physics Laboratory \& Department of Electrical and Electronic Engineering, University of Bristol, Merchant Venturers Building, Woodland Road, Bristol BS8 1UB, United Kingdom}

\author{S\"oren Wengerowsky}
\affiliation{Institute for Quantum Optics and Quantum Information - Vienna (IQOQI) \& Vienna Center for Quantum Science and Technology (VCQ), Vienna, Austria \& Institut  de  Ciencies
Fotoniques (ICFO),  Barcelona Institute  of Science and Technology, 
08860 Castelldefels, Barcelona, Spain}
\author{Martin Lon\v{c}ari\'{c}}
\affiliation{Photonics and Quantum Optics Research Unit, Center of Excellence for Advanced Materials and Sensing Devices, Ru\dj{}er Bo\v{s}kovi\'{c} Institute, Zagreb, Croatia}
\author{Sebastian P. Neumann}
\affiliation{Institute for Quantum Optics and Quantum Information - Vienna (IQOQI) \& Vienna Center for Quantum Science and Technology (VCQ), Vienna, Austria}
\author{Bo Liu}
\affiliation{College of Advanced Interdisciplinary Studies, NUDT, Changsha, 410073, China}
\author{\v{Z}eljko Samec}
\affiliation{Photonics and Quantum Optics Research Unit, Center of Excellence for Advanced Materials and Sensing Devices, Ru\dj{}er Bo\v{s}kovi\'{c} Institute, Zagreb, Croatia}
\author{Mario Stip\v{c}evi\'{c}}
\affiliation{Photonics and Quantum Optics Research Unit, Center of Excellence for Advanced Materials and Sensing Devices, Ru\dj{}er Bo\v{s}kovi\'{c} Institute, Zagreb, Croatia}
\author{Rupert Ursin}
\affiliation{Institute for Quantum Optics and Quantum Information - Vienna (IQOQI) \& Vienna Center for Quantum Science and Technology (VCQ), Vienna, Austria}
\author{Stefano Pirandola}
\affiliation{Department of Computer Science, University of York, York, YO10 5GH United Kingdom}
\author{John G. Rarity}
\affiliation{Quantum Engineering Technology Labs, H. H. Wills Physics Laboratory \& Department of Electrical and Electronic Engineering, University of Bristol, Merchant Venturers Building, Woodland Road, Bristol BS8 1UB, United Kingdom}

\author{Siddarth Koduru Joshi}
\thanks{ Correspondence and requests for materials should be addressed to
 Siddarth Koduru Joshi \href{mailto: SK.Joshi@Bristol.ac.uk}{SK.Joshi@Bristol.ac.uk}}
\affiliation{Quantum Engineering Technology Labs, H. H. Wills Physics Laboratory \& Department of Electrical and Electronic Engineering, University of Bristol, Merchant Venturers Building, Woodland Road, Bristol BS8 1UB, United Kingdom}


\begin{abstract}
The global interest in quantum networks stems from the security guaranteed by the laws of physics. Deploying quantum networks means facing the challenges of scaling up the physical hardware and, more importantly, of scaling up all other network layers and optimally utilising network resources. Here we consider two related protocols, their experimental demonstrations on an 8-user quantum network test-bed, and discuss their usefulness with the aid of example use cases. First, an authentication transfer protocol to manage a fundamental limitation of quantum communication -- the need for a pre-shared key between every pair of users linked together on the quantum network. By temporarily trusting some intermediary nodes for a short period of time (<40\,min in our network), we can generate and distribute these initial authentication keys with a very high level of security.
Second, when end users quantify their trust in intermediary nodes, our  flooding protocol can be used to improve both end-to-end communication speeds and increase security against malicious nodes.

\end{abstract}
\maketitle


\maketitle

\section{Introduction}


Quantum Key Distribution (QKD) is a point to point protocol for  communication with security based on fundamental laws of physics~\cite{Pirandola:20}. 
Recent advances in quantum networks have enabled two-party QKD protocols to interconnect an increasing number of users~\cite{wengerowsky2018entanglement, 8user, liu2020entanglement, shi2020stable,lingaraju2020adaptive}. Minimising the resources needed for such networks and optimising their utilisation are essential steps towards their real-world deployment. Several quantum networks trade security for practicality by using trusted nodes to relay the message/keys between end users~\cite{Peev2009,Sasaki2011,Stucki2011,Xu2009,FieldQKD_Wang14}, while access networks and entanglement-based networks do not rely on trusted nodes for their functionality~\cite{Toliver2003,Chen:10-chinese-access-network,Elliott2005,Chang2016,wengerowsky2018entanglement, 8user}.

There is an often overlooked cost to deploying a quantum network -- authentication. Quantum communication assumes that users share a public but authenticated classical communication channel,
which requires a pre-shared secret key. In a fully connected network, every user must maintain a secure database of pre-shared authentication keys with every other user. As quantum networks grow, this rapidly becomes impractical, to the extent that government cyber security agencies QKD~\cite{ncsc2020} highlight the algorithms used for the inaugural authentication as a major security weakness.
The ideal solution for quantum networks is to find a practical way for users to establish initial authentication keys on the fly and as needed. The ideal quantum network
solution should also have a security that is better than any possible classical or post-quantum algorithm. Furthermore, large networks need to optimally utilise the communication bandwidth of all available links to maximise network
throughput.

Naively, both the above tasks could be solved by using trusted nodes. Authentication keys between two users could be established via referral from mutually trusted nodes. The more mutually trusted nodes used for this referral process, the greater the security against any one trusted node potentially being malicious. 
Conversely, instead of using multiple independent paths to improve the final security, we could concatenate the resulting keys to boost the total end-to-end key generation rate, as is the case in a flooding protocol. 
Naturally, this is only possible when the end users treat all nodes along each path as trusted nodes. Typically, when end users use trusted nodes in quantum networks, they must then place complete trust in those nodes forever. For long term data security this is neither viable nor practical.  Is it possible for end users to place partial and/or temporary trust in intermediary nodes in a quantum network?

In this paper we present two closely linked protocols to address the above question. Firstly, we present a technique to authenticate users in a quantum network with the least number of pre-shared keys (Section~\ref{sec:SIAT}). Secondly, another technique to improve end-to-end key generation speeds based on flooding, in accordance with previous information theoretic results (Section~\ref{sec:FloodingKey}) ~\cite{pirandola2019end}. We use the definitions of trust and adversaries given in Section~\ref{sec:terminology}. In Section~\ref{sec:eg} we examine the utility of these protocols through illustrative use cases, which are demonstrated on our fully connected quantum network. Since any other network topology can be considered as a sub-graph of such a fully connected network, our demonstrations are generally applicable to quantum networks. Furthermore, Section~\ref{sec:newUser} emphasises the close relationship between these  two protocols by demonstrating their use in adding a new user to the quantum network. All the demonstrations were implemented on an 8-user entanglement-based quantum network described in Ref.~\cite{8user} and in Section~\ref{sec:Net}. We demonstrate a considerable improvement in terms of the quantum network's scalability, versatility, security and throughput.

\section{Practical limits on and types of trust in nodes} \label{sec:terminology}

Practical usage scenarios ultimately dictate the network design and protocols that operate on the network. Consider a typical scenario where various users on the network belong to multiple communities or organisations. Based on their interactions, each user in the network has different categories of trust for all other known nodes. Furthermore, there may be several users unknown to any given user. In a quantum network with information-theoretically perfect security, these considerations are paramount, and in a large network it is often more important to assign a level of trust to nodes rather than a binary trusted or untrusted state. We therefore define four categories of trust. Note that a particular node can be assigned different levels of trust from different users.

\begin{itemize}
    \item A \textbf{trusted} node (or passive eavesdropper) acts as is required of them within the protocol. They may read communication passed through them but will not change it, or broadcast it publicly (they may still access information that is broadcast publicly). 
    \item A \textbf{dishonest} node may do everything in their power to interrupt and/or intercept communication between Alice and Bob, including (but not limited to) reading any communication between them and broadcasting the message publicly (potentially without the knowledge of Alice and Bob). 
   
    \item \textbf{Partial/temporary trust}: 
    An end user can assign an intermediary node a certain probability of being trusted. This subjective estimate is called partial trust. Users provide the network with this measure based on their experience with other nodes. It is also possible to specify a time duration within which any given intermediary node can be  trusted/partially trusted. This is temporary trust.
\end{itemize}

The case of a node which is completely trusted (and therefore does not read or reveal any communication that they pass on) is trivial and therefore not considered in this work.

In general, we can consider any node or combination of nodes in the network to be an adversary. Additionally, several nodes could conspire together, in which case they can be considered a dishonest single adversary.

\begin{itemize}
    \item \textbf{A collective adversary, $c_a$}:  When $c_a$ or fewer nodes in the network are malicious with or without additional external eavesdroppers, they form a single collective adversary with a bound of $c_a$ users.
    This concept was introduced in Ref.~\cite{Salvail10}.  
\end{itemize}

\section{Secure Inaugural Authentication Transfer Protocol}
\label{sec:SIAT}

QKD offers provable security \cite{Pirandola:20}, ensuring that messages cannot be intercepted or decoded. However it requires both a quantum channel and an authenticated classical channel, which requires that the users pre-share a key. Additionally, the classical hardware or people sending the message can be compromised. A National Cyber Security Centre white paper discussing QKD highlights the dependence on authentication as a major flaw \cite{ncsc2020}.

Thus, the most practical way to deploy a single QKD link (between site $A$ and site $B$) in the field is for both QKD devices to be initiated with the same one-time authentication key. Then two trusted teams of people must accompany the clean un-compromised hardware to the installation sites $A$, $B$ and commission them. However, installing a new user into an existing quantum network of $n$ users will require one team to accompany the new hardware and $n$ teams to install new authentication keys with all the existing users.

Here we present an alternative, where deploying a new node requires substantially less effort. The inaugural authentication key is sent through a trusted intermediary node, that has a secure connection to the desired end users. To prevent a man-in-the-middle attack, the node could be monitored in person, but this is necessary only for a short period of time (as illustrated in Fig. \ref{fig:new user}). Alternatively, as discussed in Section \ref{sec: scalability}, a new user only needs inaugural authentication keys to be securely sent to 2 users in order to be able to initialise a QKD scheme with any other user of the network (assuming that the underlying network is fully connected).

\subsection{Background}

Authentication schemes are a method of producing a tag, dependent on the message and a pre-shared key. This demonstrates that the sender's message is unchanged and was sent by the correct user. All classical communication in a QKD protocol must be authenticated; the communication can be public but must not be tampered with by a malicious party \cite{BB84}.

All QKD protocols assume that the users already share authentication keys. For two parties this is trivial; the users can meet in person to verify their identities and exchange initial keys. However, in a large network, the required number of pre-shared authentication keys grows quadratically. Besides being impractical, it reduces the functionality of a future quantum internet if communication is only secure between ``known'' users.

The Wegman Carter (WC) authentication protocol \cite{Wegman1981NewEquality}, which is the most widely employed, uses hash functions. All classical communication between Alice and Bob is ``signed''  with the appropriate tag, and an adversary Eve can only replace the message with her own if she is able to guess an appropriate tag. The authentication is compromised when Eve is able to ﬁnd another message for which she is able to guess a tag, given her knowledge of a previous message-tag pair. The probability of this happening is given by $\epsilon$, which parametrises the insecurity.

When using the WC scheme it is provably impossible for an adversary to have a higher probability of success than $\epsilon$. Therefore an arbitrary level of security can be achieved, depending on the length of the initial shared key. However, reusing the same key can pose a security risk \cite{portmann}, and therefore a certain proportion of the key generated in a round of QKD is used in the authentication of further rounds. Section \ref{sec:newUser} further considers the parameters of the WC scheme in a likely experimental implementation.

\subsection{Security of authentication with one trusted node}
\label{sec:Auth}

Distributing inaugural authentication keys via a third party requires that significant trust be placed in an intermediary node. Here we demonstrate that such trust need not be permanent. Instead, we show that the intermediary may be restricted to a short window in which to perform an impostor attack. Once this opportunity has passed, the intermediary has no advantage over an arbitrary eavesdropper and standard QKD is sufficient to provide security.

Consider the following 3-party scenario. Alice and Bob have both been individually conducting QKD protocols with Chloe. Additionally, Alice and Bob are connected by a quantum channel but initially do not share a key that can be used to authenticate their classical channel. Since Alice and Bob can communicate securely with Chloe, she can be used as a trusted third party, to securely distribute an initial authentication key $k_{Auth}$. This key is used, following the WC scheme, to authenticate Alice and Bob's classical communication channel. Alice and Bob can now perform a QKD protocol to grow a new secret key $k_{ab}$.

This new key is secure against an arbitrary eavesdropper who does not have access to $k_{Auth}$. However, Chloe could use her knowledge of $k_{Auth}$ to falsify the classical communication during the QKD protocol and perform an impostor attack. If Alice and Bob were to continue using $k_{Auth}$ to authenticate their classical communication then Chloe would always retain the ability to conduct such an attack. In this case, Alice and Bob must trust Chloe for the entire duration of their communication.

Instead Alice and Bob can use their new key $k_{ab}$ to authenticate their classical communication. If Chloe can be trusted for the time taken to generate $k_{ab}$ then Chloe's knowledge of $k_{Auth}$ provides no information about $k_{ab}$. From this point on, Chloe has no advantage over an arbitrary eavesdropper. It is therefore only necessary to trust Chloe during the distribution of $k_{Auth}$ and for the time it takes for Alice and Bob to generate $k_{ab}$. After this point, even if Chloe were to become malicious, Alice and Bob's communication remains secure.

This can be used to communicate the initial key using the Secure Inaugural Authentication Transfer Protocol (SIAT):
\begin{enumerate}
    \item A trusted node (Chloe) sends the authentication key to two users wishing to communicate (Alice and Bob);
    \item The shared key is used to authenticate Alice and Bob's channel - once this round of QKD has finished, trust in Chloe ends (the length of the QKD round can be adjusted appropriately based on how long this is considered to be possible);
    \item The key produced in this round is used to authenticate further rounds of QKD.
\end{enumerate}
An important aspect of this protocol is that it is decentralised, allowing users to build connections without the intervention of a network authority (which is of benefit compared to previously suggested methods \cite{liu2020experimental}). Nevertheless, this can be combined with a network authority (using this as the trusted node in every case) if required. However, in the case of this protocol and the protocols discussed in further sections, the topology of the network must be known.

Finally, Ref.~\cite{Cederlof08} describes the security of authentication given partial knowledge of the key. This is not unlikely due to data leakage, and could happen if Chloe's data storage is partially compromised. This could potentially lead to an attack in which Eve is able to break the authentication system. However, until she has gathered enough information on the key to be able to falsify messages with a low probability of being detected, this does not lead to knowledge of the key $k_{AB}$ produced. Therefore, if the length of each round is sufficiently short, Alice and Bob are able to generate a new authentication key before Eve is able to eavesdrop. Furthermore, Ref.~\cite{Cederlof08} describes several possible preventions of an attack. Therefore, it is reasonable to assume that WC authentication has the security described.

\subsection{Multiple trusted nodes} \label{sec:multiple nodes}

Consider the case in which, instead of a single intermediate node who is used to share the initial authentication key between Alice and Bob, there are $n$ distinct paths (of any length) between them, and Alice can send a bit string $k_i$ through each node, with $k_{\text{Auth}} = k_1 \oplus k_2 \oplus... \oplus k_n$, where $\oplus$ is the bitwise sum modulo 2. As shown in Ref. \cite{Salvail10}, if any number $m<n$ are malicious (but do not publicise their part of the key), they still do not have access to the authentication key, as each $k_i$ provides no information on the total key. If all parties collude, they act as a single malicious party with knowledge of the full $k_{Auth}$, as above.

Ref. \cite{Salvail10} further considers the case of this network being corrupted by collective adversaries, as defined in Sec. \ref{sec:terminology}, in which up to $c_a$ nodes conspire and act as a a single common adversary. If the corrupted nodes do not publicly publish their keys (so keep the key to themselves), it is shown that $c_a+1$ non-overlapping paths between users guarantees authenticity (there is guaranteed delivery of unchanged classical messages or a notification of failure). Thus, $c_a+2$ disjoint paths are needed in the case that the corrupted nodes are dishonest and broadcast the communication publicly. 

This protocol would require active use of all of the nodes in the MNOPs. However, this does not necessarily mean that every node in the network must be used. Any node that is unavailable, for example due to being in use for other operations or being considered completely untrustworthy, can be discounted from the MNOPs as long as an appropriate topology can be formed without them.

\subsection{Practicality of partially trusted nodes}\label{sec:pract partial trust}

\begin{figure}
    \centering
    \includegraphics[width=0.8\linewidth]{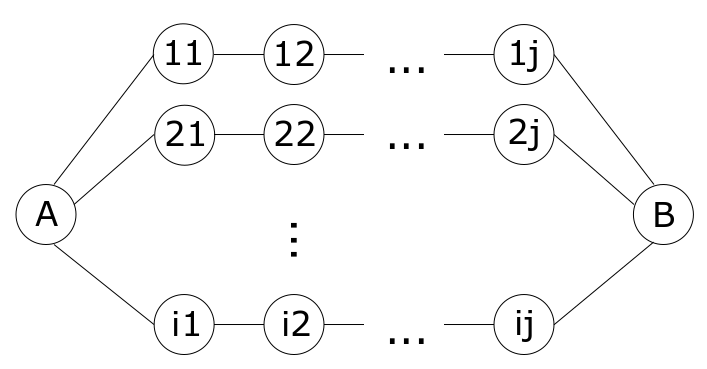}
    \caption{\textit{Example of MNOPs}. Demonstrates the labelling of the nodes in multiple non-overlapping paths (MNOPs) between $A$ and $B$.}
    \label{fig: MNOPs}
\end{figure}

Trusted nodes are a security risk but their advantages can often outweigh these concerns. Nevertheless, it is impossible to be 100\,\% certain about an intermediary node. If each node is assigned a risk factor then we can use multiple non-overlapping paths (MNOPs) to mitigate this risk. If some probability of insecurity can be tolerated, the insecurity of the initial key is as follows.

Let every node be characterised (in agreement between Alice and Bob, or taking the lowest trust value for each) by the partial trust probabilities $T_j$ (as defined in Section \ref{sec:terminology}). That is, they are honest with probability $T_j$, although they may read any information to which they have access. With probability $1 - T_j$, they are completely dishonest and may share information publicly. Consider an initial key $k_{\text{Auth}}$ to be shared between users Alice and Bob, using several disjoint paths that are labelled $i$, each with nodes labelled $j$, as shown in Fig. \ref{fig: MNOPs} (similarly to the multiple paths shown in Ref. \cite{Salvail10}). The probability that the key is compromised (can be read by a node) is:
\begin{equation} \label{compromised}
   P_{comp} = \prod_i \biggl( 1 - \prod_j T^{ij} \biggr) + \sum_k \bigg( \prod_{j'} T^{k,j'} \prod_{i, i \neq k} \biggl( 1 - \prod_j T^{ij} \biggr)\bigg) ,
\end{equation}

where $k$ runs over the set of $j$. The first term represents the scenario that at least one node in every path is dishonest (equivalent to the key in that path being published), and the second term describes the scenario in which at least one node in every path but one is dishonest (meaning the nodes in the remaining path can infer the key).

There is also the case in which the paths overlap, and cross at a particular node. When considering a collective adversary bounded by $c_a$, this effectively reduces the number of paths. Applied to Eq. \ref{compromised}, the cross-point node can be considered as occupying a position in multiple paths (say, $i$ and $i'$) so, for example, $T^{ij}=T^{i'j}=T^{xx}$ where $xx$ labels the cross-point. Additionally, we must account for the case in which the cross point node collates the information from several otherwise trusted channels, and at least one node on all other channels is dishonest. This modifies Eq. \ref{compromised} by adding an additional term $T^{xx} \prod_{\{i\}/x} \left( 1 - \prod_j T^{ij} \right)$ for each cross-point.

\subsection{Scalability of initial key distribution in a quantum network} \label{sec: scalability}

Being able to verify the identity of users is necessary to build a secure quantum network. Therefore, whenever a new user joins a network, they must physically receive a secret key ($k_{\text{Auth}}$) that they share with all other users in the network. Thus for a network of $n$ users, this must be done at least $n-1$ times. However, for the new user to be able to have authenticated communication with all of the other users of the network, they need to share a secure initial key with each desired end user, and so that number increases depending on the trust in different users, the available resources, and the network topology. As this may require a considerable amount of cost and effort, the number of pre-shared keys should be minimised. 

Consider the scenario where the user being added to the network has full trust in all of the members of the network. Two disjoint paths are needed between the users wishing to communicate, as discussed in Section \ref{sec:multiple nodes}. This prevents intermediate nodes having access to $k_{\text{Auth}}$ if they become malicious. In the case of a fully connected network with $n>2$, this means that two keys need to be distributed for each new user (as there will be at least 2 disjoint paths to every other user), so the number of pre-shared keys required increases as $2n-3$.

As discussed previously, in the case where nodes collude or are corrupted by a collective adversary of $c_a$ nodes, there must be $c_a+2$ disjoint paths between two users wishing to communicate to distribute the shared key (with the other nodes being trusted). Within a fully connected graph where each user wishes to communicate with every other user, the number of pre-shared keys ($n_k$) therefore scales as \begin{equation} \label{eq: scaling} n_k = \frac{(c_a+2)(c_a+1)}{2} + (n-c_a-2)(c_a+2)\end{equation} for $n>c_a+1$. Given that $c_a$ scales sub-linearly in $n$, $n_k$ is linear in $n$, in comparison to the naive solution, which would scale quadratically in $n$.

A full derivation of this equation is given in Supplementary Information, Section \ref{sec: scaling deriv}. 

For other topologies, this number will depend on the availability and length of MNOPs that can be found in the network. However, any other topology can be considered a subnetwork of a fully connected network. We assume that the appropriate physical infrastructure exists, so that a fully connected network could first be created with the given number of preshared keys. Then, for any other topology, only some of these secure connections, once established, would need to be used. Therefore, Eq. \ref{eq: scaling} can be considered an upper bound for any topology.

This only considers the situation in which there are sufficient physical connections to allow this - for more complex topologies in both the physical infrastructure and communications requirements, futher calculation would be necessary, however the given principles (namely Eq. \ref{eq:reduced trust} and the SIAT protocol) still hold.

\section{Optimal key rates using key flooding with trusted nodes}
\label{sec:Flood_Trusted}
\subsection{Background}
\label{key flooding}
Flooding is a classical routing protocol for transmission of information through a multi-hop network \cite{DBLP:books/lib/TanenbaumW11}. This strategy was recently employed in quantum information theory to show that multi-path quantum (and private) capacities of a quantum network can greatly outperform corresponding single-path capacities~\cite{pirandola2019end}. When a flooding protocol is used in the communication between two users of a network, the source broadcasts a data packet to every user to which it is connected. Each intermediary node then manipulates and outputs the incoming packets on every possible outgoing link except the one(s) it arrived from. This continues for every node on the network except the receiver. In this way every link in the network is used exactly once and multiple paths through the network are used in parallel.

Such a protocol has a number of benefits. Each intermediary node does not need to know the full topology of the network, merely the nodes with which they share links. Additionally, flooding protocols are very robust. As long as at least a single path exists between the two end users, communication between them will occur. Flooding is therefore a general protocol which may be applied in any network and we make no assumption on the network topology in describing the protocol, except that it is known.

\begin{figure}
    \centering
    \includegraphics[width=0.95\linewidth]{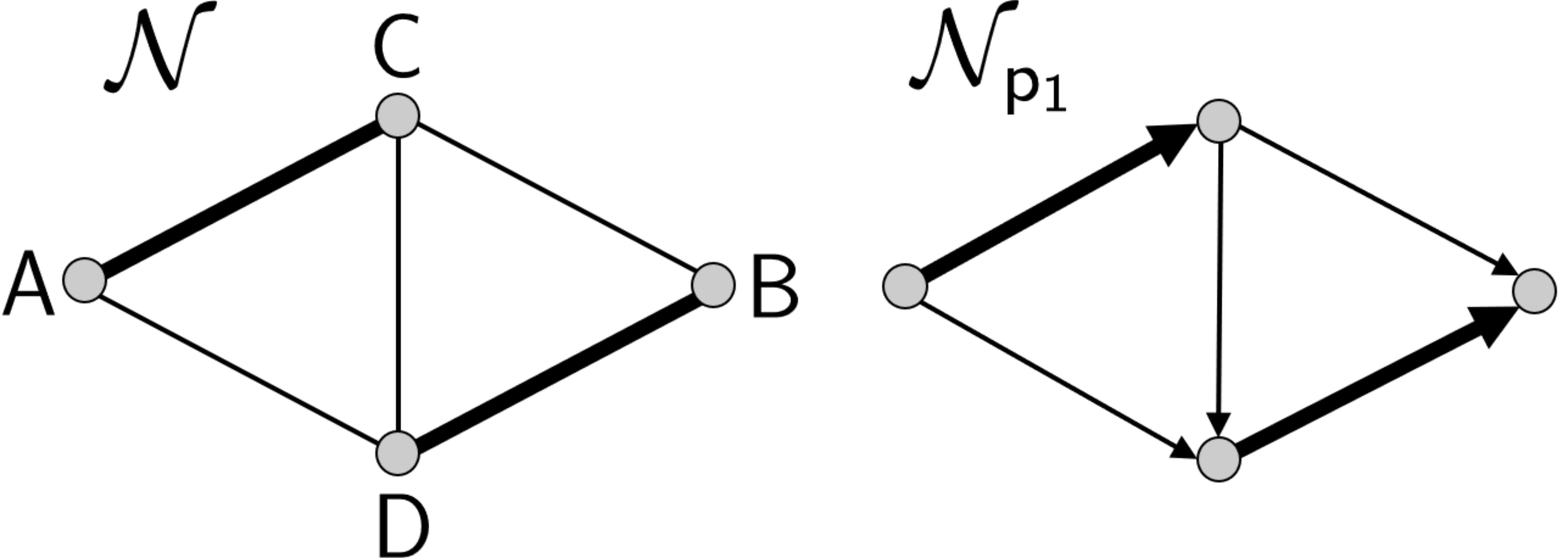}
    \caption{\textit{Comparison of the simple graph representing the network and the directed graph representing a flooding protocol performed on it}. The left image depicts the graph of the 4 user network $\mathcal{N}$ comprised of end users $A$, $B$ and two intermediary nodes $C$, $D$. $A$ seeks to communicate with $B$ via a flooding protocol. The thickened lines represent higher throughput connections between $A$ and $C$ and $B$ and $D$. The first flooding protocol $\boldsymbol{p_{1}}$ converts this graph into the directed graph $\mathcal{N}_{p_{i}}$ which is depicted on the right. The second possible flooding protocol $\boldsymbol{p_{2}}$ simply corresponds to reversing the orientation of the edge connecting $C$ and $D$.}
    \label{directed}
\end{figure}

In general to implement flooding in a given network, there exists more than one routing strategy.  This is best seen when network is represented by a simple graph
$\mathcal{N}(V,E)$ where the set of vertices $V$ represents the users of the network and the set $E$ represents the edges. Two vertices $\boldsymbol{v_{i}}$ and $\boldsymbol{v_{j}}$ are connected by an edge $(\boldsymbol{v_{i}},\boldsymbol{v_{j}}) \in E$ if the corresponding users share a connection. Flooding uses every edge in the network exactly once, so a given flooding routing strategy $\boldsymbol{p_{i}}$ corresponds to assigning an orientation to every edge in the network which represents the direction of information flow. For the case of two users communicating, in which one acts as a source and the other a sink, there is the additional requirement that all the edges from the source are orientated positively and the edges into the sink are orientated negatively, as typical in a flow network. Orientating the edges transforms the simple graph $\mathcal{N}$ into a directed graph $\mathcal{N}_{p_{i}}$ which represents the $i$th possible flooding routing strategy on the network. This is depicted for a 4 user `diamond' network in Fig.~\ref{directed}. Since each possible directed graph corresponds to a routing strategy and each intermediary edge can be orientated in one of two possible directions, there are $2^{(|E|-|E_{\mathrm{s}}|)}$ possible routing strategies for flooding where $|E|$ is the total number of edges and  $|E_{\mathrm{s}}|$ is the number of edges connected to the source or sink. 

In the case of a QKD network, some adaptations must be made. Since we use keys shared between users in the network as one-time pads, there is clearly a maximum amount of information that may be securely transmitted between two users. Therefore, rather than broadcast the full amount of received information, an intermediary node only outputs as much as may be communicated securely to each of its neighbouring nodes. Nonetheless we make use of each edge of the network exactly once and the intermediary users only require knowledge of which users they share keys with and the lengths of those keys. 

This strategy of quantum secure flooding is a purely-cryptographic formulation of the flooding protocol designed in Ref.~\cite{pirandola2019end} to lower-bound the multi-path quantum and private capacity of a quantum network. The idea is that, once entanglement or secret bits are shared between the nodes of a quantum network, the operations of
entanglement swapping or key composition (one-time pad) can be combined with an optimal multi-path routing strategy~\cite{pirandola2019end}. As we can expect from a multi-path protocol, quantum secure flooding provides advantages over single-path strategies for communication on a QKD network. In the case that the intermediary nodes are fully trusted, flooding may be used to increase the end-to-end key rate, as we demonstrate experimentally in Section \ref{sec:FloodingKey}. In the case that the intermediary nodes are only partially trusted, flooding may instead be used to reduce the risk associated with using these nodes. We demonstrate the latter scenario in Section \ref{sec:FloodingSecurity}.
\subsection{Linear Chain}
\label{sec:linear chain}

Any multi-path protocol may be considered as multiple single path routing protocols taking place simultaneously. It is therefore natural to first consider how two users may share a key using a single path protocol.

\begin{figure}
    \centering
    \includegraphics[width=\linewidth]{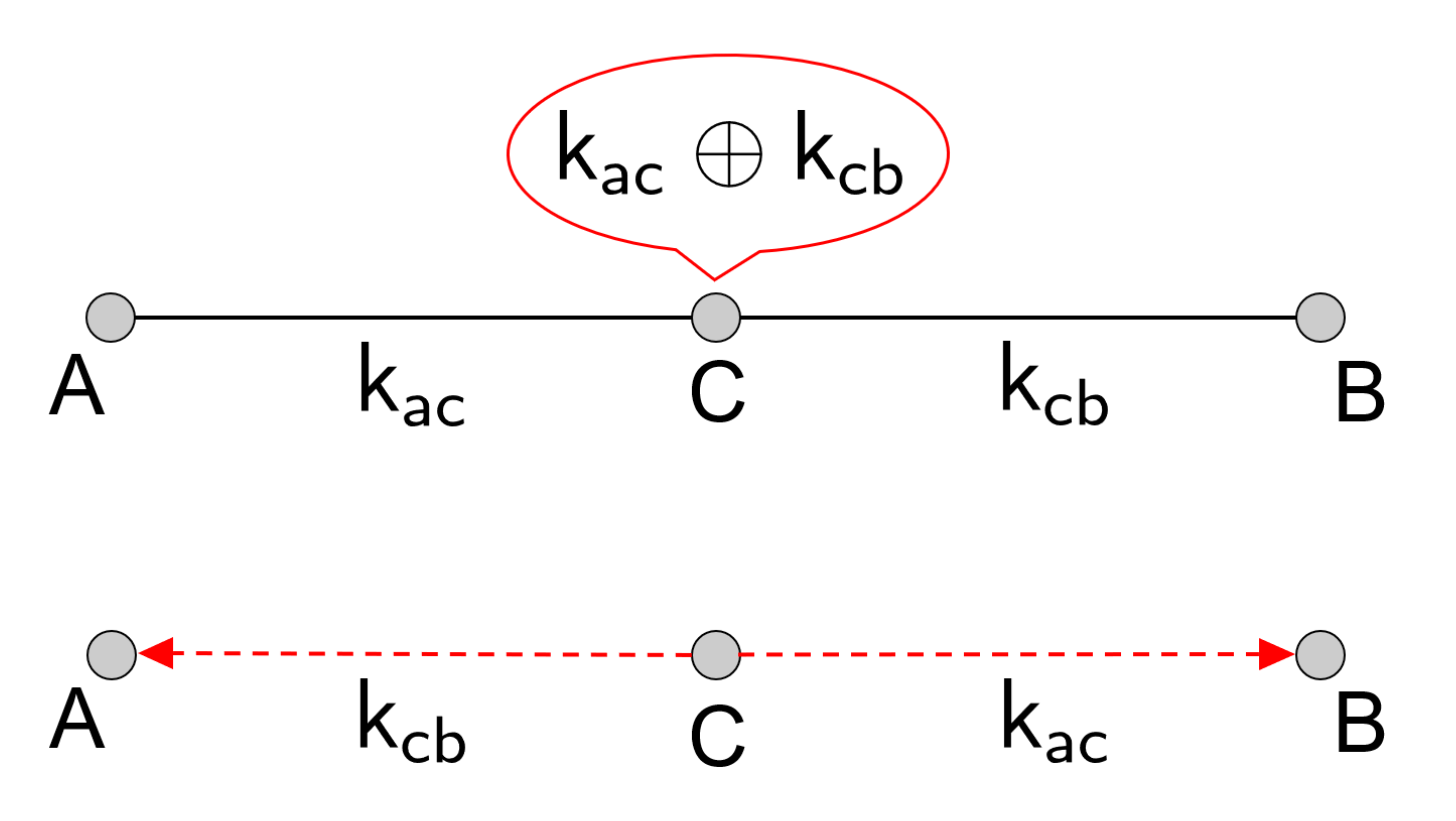}
    \caption{\textit{Simple key flow}: The top shows the intermediary node $C$ announcing the combined key $k_{ac} \oplus k_{cb}$ which is shown in the top diagram. $A$ and $B$ may decode the keys $k_{cb}$ and $k_{ac}$ respectively. Therefore knowledge of $k_{cb}$ ($k_{ac}$) flows to $A$ ($B$) via $k_{ac}$ ($k_{cb}$). This is depicted in the bottom.}
    \label{3 person}
\end{figure}

The simplest network with a single path between two users is a 3-user chain network. We consider as an example, three users Alice ($A$), Bob ($B$) and a trusted intermediary node ($C$), in which Alice and Bob do not directly share a key, but instead both share a key with the intermediary node. For simplicity we consider the length of the keys $k_{ac}$ and $k_{cb}$ to be equal. The intermediary node makes a public announcement of the bit-wise sum modulo 2 of the two keys $k_{ac} \oplus k_{cb}$. Alice and Bob can decode the other's key by again performing bit-wise addition of their own key with the announced combined key. Knowledge of $k_{ac}$ flows from $C$ to Bob via $k_{cb}$ and correspondingly knowledge of $k_{cb}$ is passed to Alice via $k_{ac}$. This is illustrated in Fig. \ref{3 person}. 

We now make the following observations. Provided the intermediary node is trusted, the announcement is made correctly and an adversary only has access to the combined key $k_{ac} \oplus k_{cb}$ which provides no information about either of the individual keys. However, Alice and Bob may not concatenate the two keys to generate a shared key of twice the length, as the announcement provides sufficient information to determine one key from the other. It is also clear that if the original keys were of different lengths that the maximum length of secure key that Alice and Bob may share is equal to the length of the shorter key.

The natural extension to $n$ trusted intermediary nodes also reduces to the above example. Each node learns all the keys used in announcements along the chain and a general adversary only has access to the bit-wise sum of any of these keys. Thus, whilst Alice and Bob learn all the keys along the chain and these keys remain secure, they can only use one key which must be pre-determined. Therefore the maximum length of secure key that can be generated between the two end users is equal to the length of the shortest key in the chain, and corrupting a single user gives an adversary access to all the keys in that chain. Thus, given partial trust in each of the intermediaries, more users increases the probability of insecurity. We can instead increase the security and rate of the protocol by introducing multi-path strategies in more complex networks, as we now detail.

\subsection{Quantum Secure Flooding}
\label{sec:flood_protocol}
The quantum secure flooding protocol is the same regardless of whether the end users wish to increase their end-to-end communication rate or the security of their communication. In both cases, the end users share multiple secure keys via multiple paths (i.e. via different sets of other partially or fully trusted nodes). The difference arises at the end of the protocol when the end users privately either concatenate keys (in order to increase the rate) or XOR  keys (to increase security).

These scenarios are discussed in Section \ref{sec:FloodingKey} and Section \ref{sec:FloodingSecurity} respectively. Throughout this section we assume that the communication is between the end users Alice -- the source and Bob -- the sink. It is important to note that, in this protocol, it is necessary for Alice to have complete knowledge of the current network topology, and available/unused key rates between users.

As discussed in Section \ref{sec:linear chain}, keys are passed through the network by announcing them XORed with other network keys. Since a key may only be used once as one-time pad, an intermediary node should not simply pass all of their received keys out to every other node they share a connection with. Instead an intermediary node must split its received keys only among the users with which it wishes to communicate. 

\begin{figure}[tb]
    \centering
    \includegraphics[width=0.95\columnwidth]{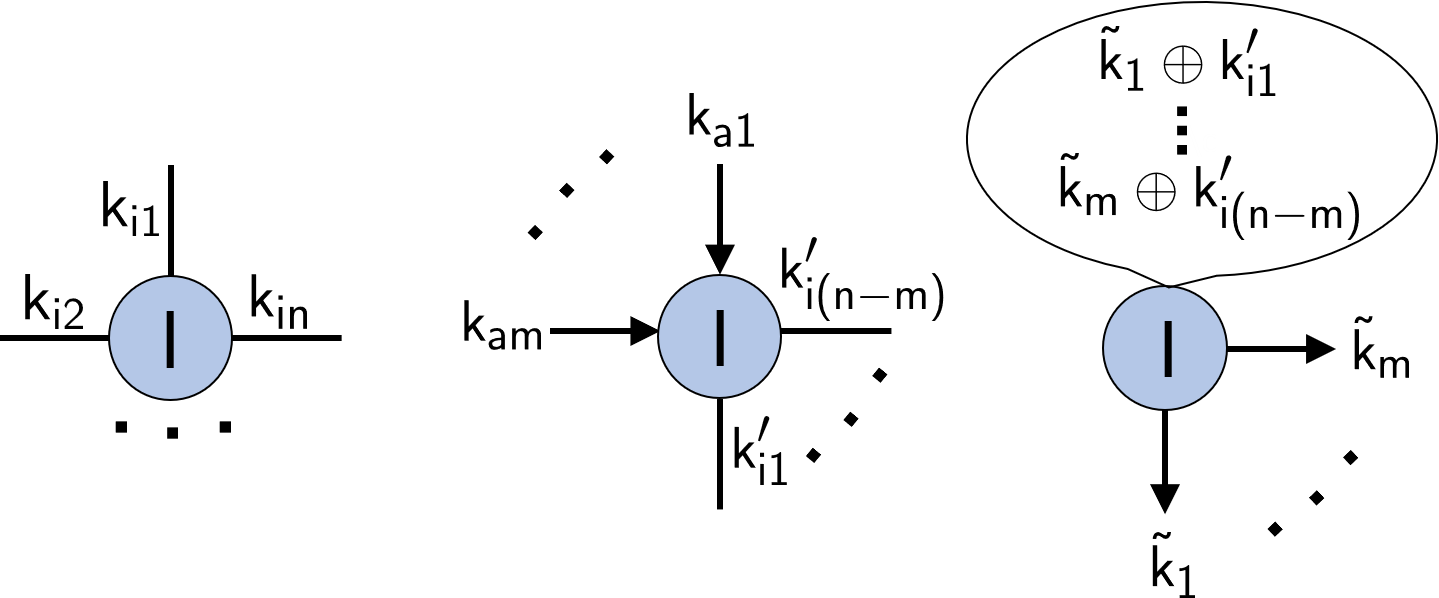}
    \caption{\textit{Procedure for splitting keys during quantum secure flooding.} The left diagram depicts the initial neighbourhood of user $I$ which shares keys with $n$ other users. The middle diagram shows the user when they are contacted during stage 4 of the quantum secure flooding protocol. The user receives $m$ keys and has $n-m$ remaining keys. The right diagram shows the node having split their keys according to the received ordering and announcing these keys XORed with the remaining $n-m$ keys. Since all of the remaining keys are used this corresponds to the case in which the total length of keys received is greater than the length of the remaining keys.}
    \label{fig:intermediary}
\end{figure}

To illustrate this, we consider as an example an intermediary node $I$ who shares keys with $n$ other nodes as illustrated in Fig. \ref{fig:intermediary}. The intermediary receives a set of $m$ keys $\{k_{a1},...k_{am}\}$ with corresponding lengths $\{l_{i1},...l_{im}\}$, either directly from the source or decoded from another intermediary's announcement. The intermediary therefore has $n-m$ keys that have not yet been used nor are keys shared with the source. The intermediary node privately concatenates all of their received keys into the combined key $\tilde{k}=k_{a1} || ... k_{am} \ $. The node also receives an ordering for their remaining $n-m$ keys, resulting in the ordered set $\{ k'_{i1},...k'_{i(n-m)}\}$. Each of these keys also has a corresponding length $\{l'_{i1},...l'_{i(n-m)}\}$. The intermediary node then splits $\tilde{k}$ into separate keys $\tilde{k}_{i}$ with lengths equal to $l'_{i}$ until the entire key is used or all the $m$ unused keys have matching length keys. $I$ then announces each of these keys XORed with the corresponding unused key $k'$.

\begin{figure}[tb]
    \centering
    \includegraphics[width=\columnwidth]{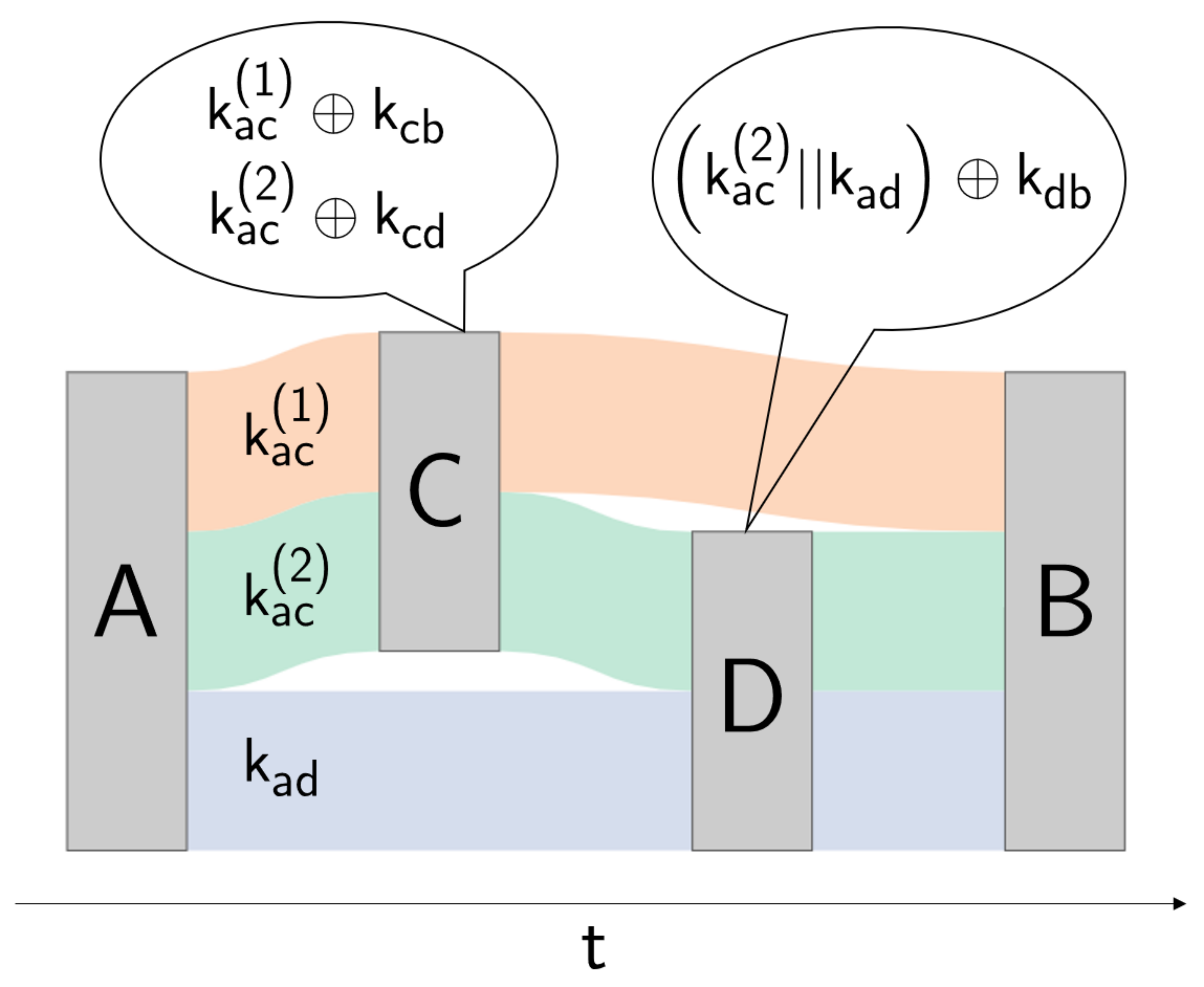}
    \caption{\textit{Key flow during flooding}:  The diagram depicts the time at which announcements are made and the flow of knowledge of $A$'s keys through the network depicted in Fig.~\ref{directed}. Each colour depicts the flow of knowledge of one of $A$'s keys (after splitting). Node $C$ splits their longer key $k_{ac}$ into two keys of equal length $k_{ac}^{(1)}$ and $k_{ac}^{(2)}$. They then publicly announce $k_{ac}^{(1)}\oplus k_{cb}$ and $k_{ac}^{(2)}\oplus k_{cd}$ causing $k_{ac}^{(1)}$ to flow to $B$ and  $k_{ac}^{(2)}$ to flow to node $D$. Node $D$ concatenates $k_{ac}^{(2)}$ and $k_{ad}$ and announces $(k_{ac}^{(2)}||k_{ad})\oplus k_{db}$ causing  $k_{ac}^{(2)}$ and $k_{ad}$ to flow to $B$.}
    \label{sank flood}
\end{figure}

In more detail, the steps of the protocol are the following:

\begin{enumerate}
    \item The optimal flooding protocol is calculated by Alice as detailed in \ref{sec:floodrouting}. In general this can be achieved in $\mathcal{O}(|V||E|)$ time, for a network with $|V|$ nodes and $|E|$ edges. \cite{Orlin2013}.
    \item Alice contacts the first intermediary node, sending them an ordered list of output keys.
    \item The intermediary user splits the keys they share with Alice according to the ordered list they received. They announce these keys XORed with their remaining keys in accordance with the received ordering.
    \item Alice contacts the next intermediary node and sends them an ordered list of output keys. They privately decode any keys they can from previous announcements. They concatenate these keys with any keys they shared directly with Alice and then split the resulting key in accordance with the ordered list they received. These keys are announced XORed with their remaining keys in accordance with the  received ordering. 
    \item Stage 4 is repeated for all the required intermediary nodes in the protocol.
\end{enumerate}
Fig. \ref{sank flood} illustrates the optimal quantum secure flooding protocol applied to an idealised network shown in Fig \ref{directed}.

Calculation of the optimal flooding protocol  requires full knowledge of the network topology. It may be possible for malicious users to convince others that actually overlapping paths do not overlap. We restrict our work here to the case in which the topology can always be determined. We note that all users in our network share entanglement, thus they can distinguish between a direct link with another user as opposed to a link via a trusted node. This may enable users to cooperatively and securely map out the network topology despite the presence of some  malicious users.

\section{Practical applications of SIAT and key flooding }
\label{sec:eg}

\subsection{Adding a new user to the quantum network}
\label{sec:newUser}
To show the feasibility of the SIAT protocol for adding a new user, we demonstrate it applied to the example of a previous experiment of an 8-user quantum network, with the data shown in Ref. \cite{8user}. 

The amount of classical communication required in order to produce a bit of key varies between experiments (as opposed to the factor of 2 in ideal QKD). In this experiment, there were up to 10 000 detection events for each bit of secure key, which were labelled with 64 bits of time tagging and 2 bits for the basis choice. This was increased to 72 bits to label each event when considering error correction data, so an estimate for the amount of classical communication required is 720 000 bits per bit of key. For an insecurity of $10^{-9}$, with 10\% of the key reused each round for authentication, this means successive QKD rounds should each produce 50 563 bits, as shown in Supplementary Info Section \ref{sec:adding user}. 

\begin{figure}[tb!]
    \centering
    \adjustbox{trim=0.1cm 0cm 0cm 0cm, clip}{
    {\includegraphics[width=1.05\linewidth]{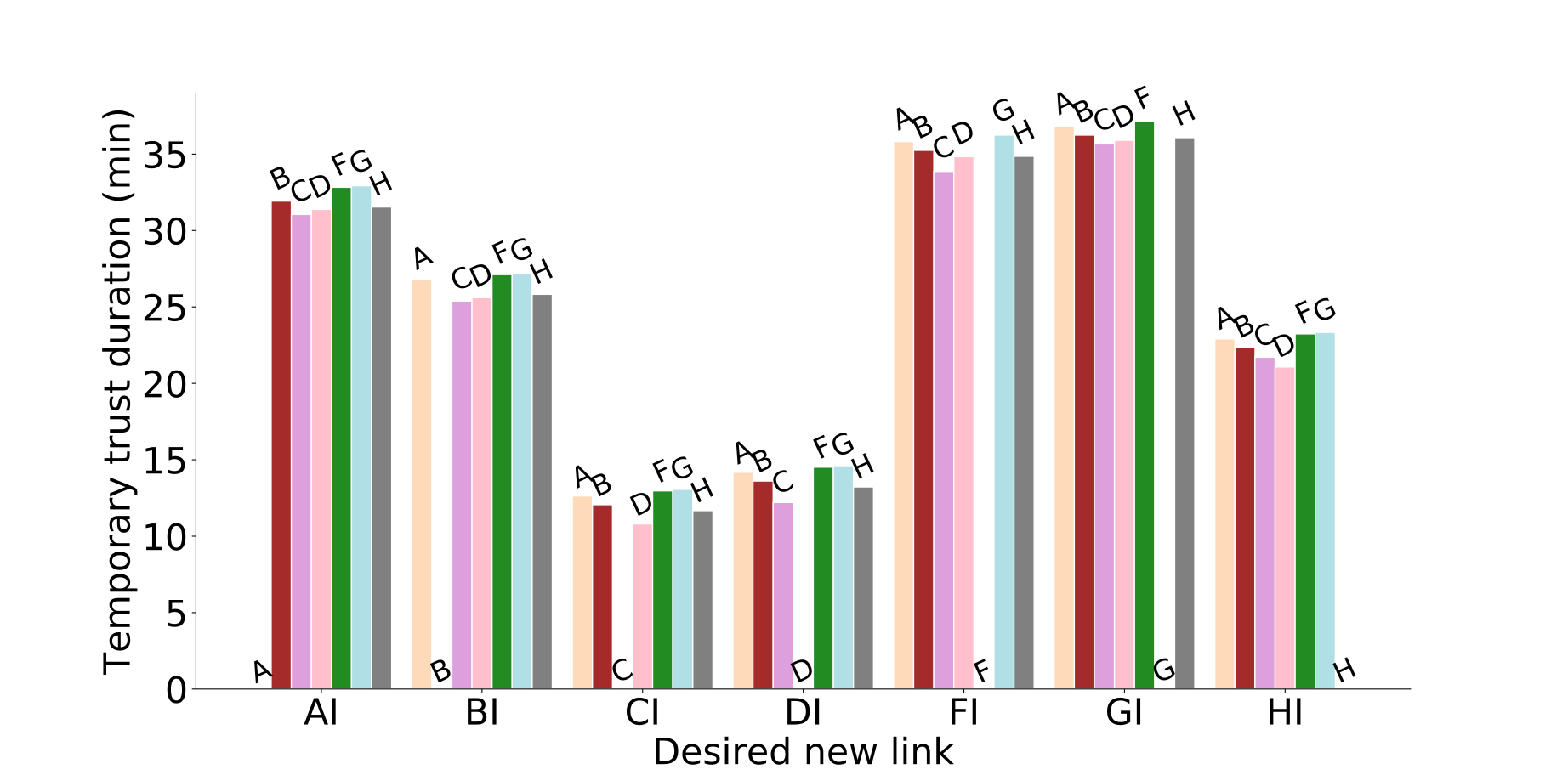}}}
    \caption{\textit{We show that initial authentication, while adding a new user to the network, can be established in less than 40 minutes.} The figure corresponds to a demonstration in which a fully connected network of 7 users exists, and a new user -- $I$ -- wishes to join the network using the SIAT protocol. It is shown the time any of the other users must be trusted as an intermediate node (users denoted by different colours and labelled above the bars) in order for $I$ to have a connection with a different desired end user (users on the $x$-axis). Data used is taken from Ref. \cite{8user}.}
    \label{fig:new user}
\end{figure}

The example case is considered in which a new user is added to the network, and wishes to form a secure connection with each other user of the network. In this example, this will be considered to be the 8th user ($I$). If each of the users ($A$, $B$, $C$, $D$, $F$, $G$, $H$) is used as a trusted intermediate node, Fig. \ref{fig:new user} shows the amount of time needed to initialise a secure, authenticated channel with any other user. This shows that, given the simplest scheme, the maximum amount of time a user needs to be trusted is less than 40 minutes. Therefore the length of time needed to initialise the authentication in a quantum network should not be considered a significant disadvantage.

However, this scheme can be improved further by utilising the previously discussed flooding protocol.

\subsection{Optimal key generation using trusted nodes}
\label{sec:FloodingKey}
\begin{figure}[tb]
    \centering
    \includegraphics[width=0.95\linewidth]{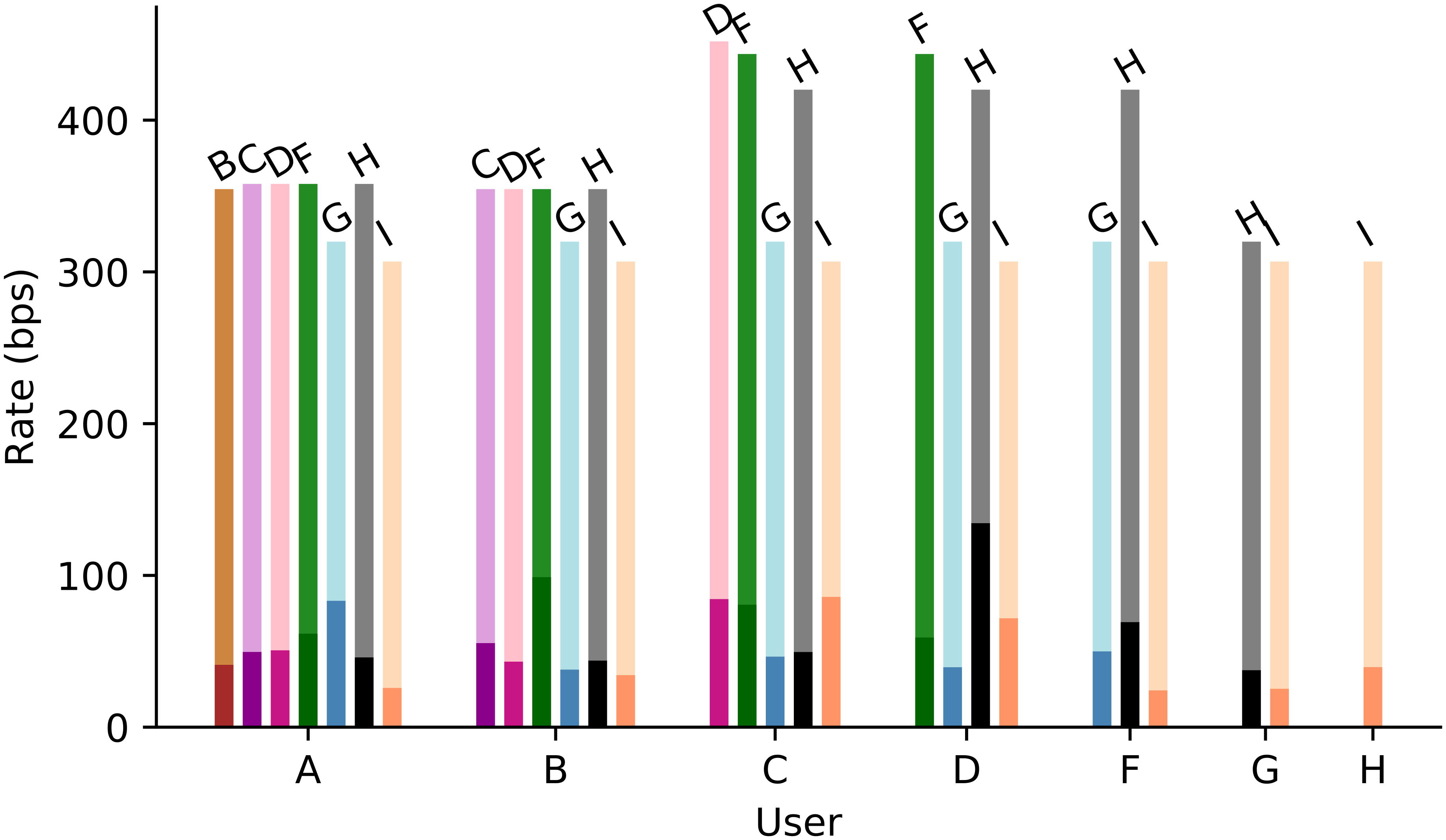}
    \caption{\textit{Improved secure key rates with end-to-end key flooding on our 8-user quantum network test bed.} The direct key rates (bold colours) and the flooding  secure key rates (lighter colours) are shown between every pair of end users in the experiment (labels above the bar indicate the user connected to).  Using the flooding protocol, a fully connected network can be made exactly equivalent to an access network with improved rates where an optical switch chooses which end users can communicate at any instant. }
    \label{optimal w trusted}
\end{figure}

\begin{figure}[tb]
    \centering
    \includegraphics[width=\linewidth]{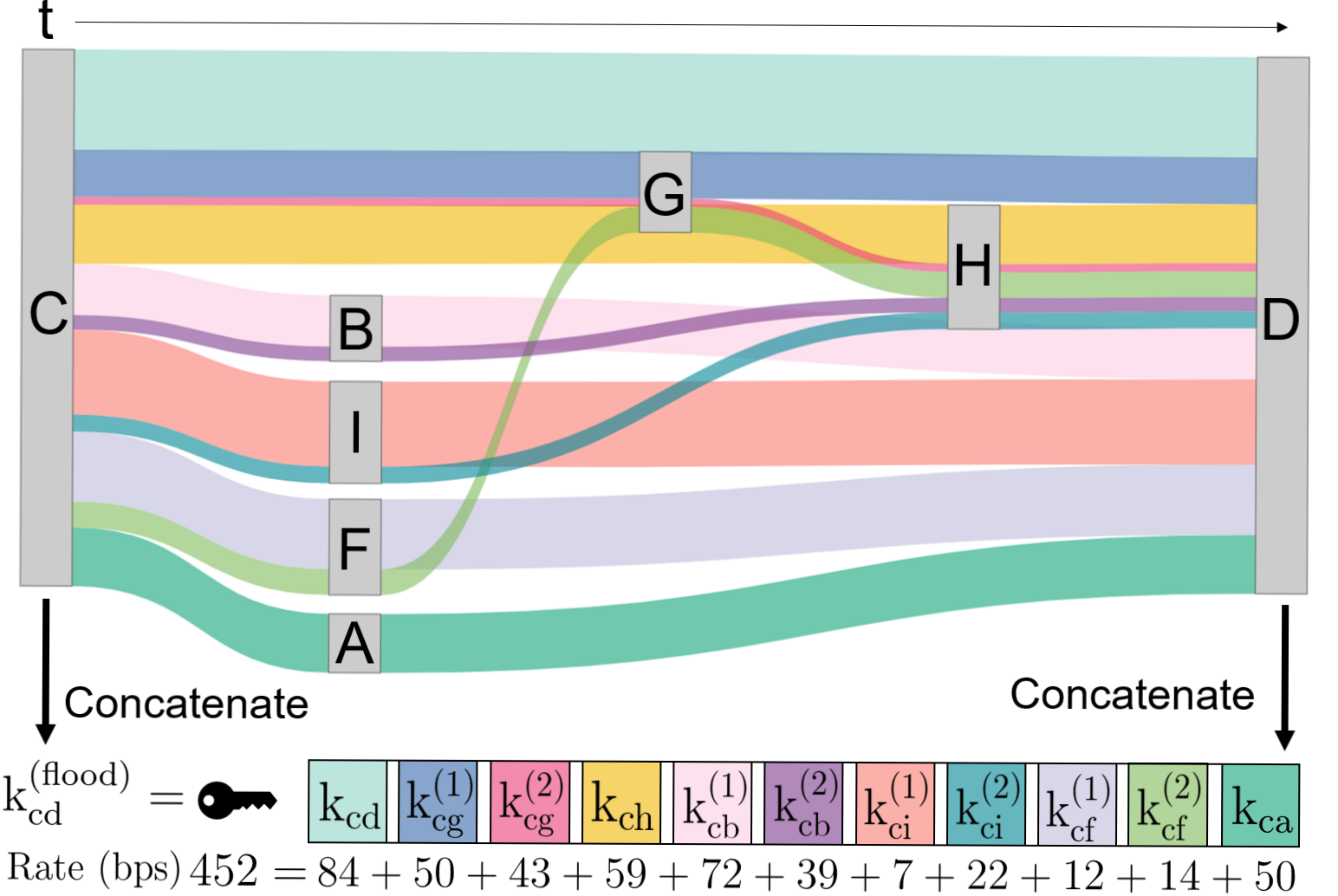}    
    \caption{\textit{Optimum end-to-end key flooding protocol between end users $C$ and $D$ on our 8-user quantum network test bed.} The figure demonstrates an optimal flooding protocol between users $C$ and $D$ corresponding to the data used in Fig \ref{optimal w trusted}. For clarity, the public announcements made by the intermediary users are suppressed and shown instead in Appendix \ref{appendix:comic}. The width of bars indicates the length of keys ultimately shared between $C$ and $D$. The end users privately concatenate the shared keys to produce $k_{cd}^{\mathrm{(flood)}}$ with length equal to the sum of the lengths of the shared keys. This achieves the optimal flooding rate shown in Fig \ref{optimal w trusted}. }
    \label{sankey trusted}
\end{figure}

The flooding protocol described in Section \ref{sec:flood_protocol} can be used to increase the end-to-end secure key rate. In the case that all the intermediary nodes are trusted (which means we are not restricted to non-overlapping paths) the two end users may concatenate all the flooded keys into a new longer key. 

We use data from the quantum network test bed described in Section~\ref{sec:Net} to demonstrate the increase in end-to-end key rate between all possible pairs of users. Data from all users is collected every 20 minutes and the entire QKD post-processing is performed including error correction and privacy amplification. 
Flooding is performed using the final secure keys. We note that it may also be possible to perform flooding on the raw or sifted keys before privacy amplification. This may result in a higher end-to-end key rate due to the non-linearity of key rate with QBER. Further work is required to consider all the security implications and study the potential improvements. Such a study should also account for  inefficiencies during the implementation of error correction and privacy amplification. 
Figure \ref{optimal w trusted} shows the optimal flooding rate for each possible pair of end users compared with the direct rate between them for a single 20 minute block. On average we were able to demonstrate an $\approx$ 7-fold increase in the key rate on our quantum network test bed. Fig. \ref{sankey trusted} shows the corresponding optimal flooding protocol between the users Chloe ($C$) and Dave ($D$). Further details are provided in Supplementary Info \ref{appendix:comic}.

\subsection{Enhancing security using flooding}
In an arbitrary network suppose there are MNOPs between two end users A,B; then the end users can choose to use the flooding protocol to maximise the secure key rate. However, this is not the only optimisation they can perform. Here we show that flooding can also be used to improve the security between the end users. 

As discussed in Section \ref{sec:linear chain}, a path through a network can be treated as a linear chain. All of the intermediary nodes in the chain learn the secure key that is passed from one end user to the other. The communication is secure along the chain as long as all the intermediary nodes are trusted. If we now consider a path $i$ with $N$ such intermediary nodes, each partially trusted with a trust value $T_j$, then the probability that the communication along the chain is secure is $T_i=\prod_{j}^{N}T_{j}$. Using single path-routing strategies the maximum security that can be achieved is simply the security of the most secure single path: $t=\mathrm{max}_{\forall i'}T_{i'}$.

However, the two end users may improve the security of their final key by first performing a flooding protocol and receiving a set of $n$ keys $\{ k_{1},k_{2},...k_{n}\}$. Each of these keys corresponds to a path through the network used in the flooding protocol and may be assigned a trust as described above. The two end users may now privately XOR all of these keys resulting in the final key  $k_{ab}=k_{1}\oplus k_{2}\oplus ... k_{n}$. In the case that all of the paths used by the flooding protocol are non-overlapping the overall trust value $t$ can be derived from Eq. \ref{compromised}.
and is given by:
\begin{equation}
t=1-\bigg(\prod_{i}^{n}(1-T_{i})+\sum_{k}T_{k}\prod_{i \ i\neq k}^{n}(1-T_{i})\bigg).
\label{trust non overlap}
\end{equation}
This follows immediately from the fact that for non-overlapping paths the protocol is compromised if either all or all but one of the paths contains a dishonest intermediary node. 
The more general case in which some of the paths overlap is also discussed in Section \ref{sec:pract partial trust}. 

\begin{figure}[tb]
    \centering
    \includegraphics[width=\columnwidth]{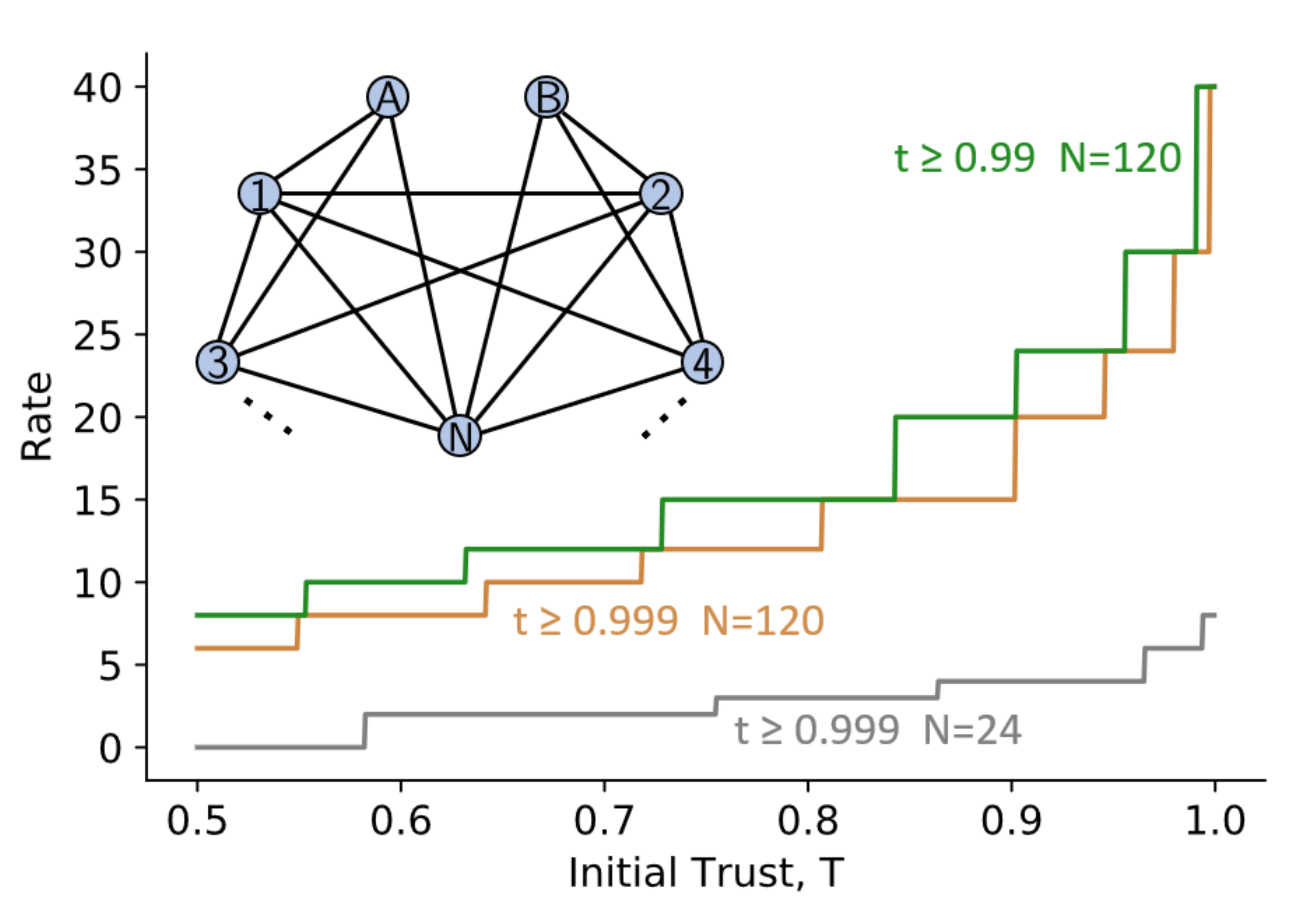}
    \caption[\linewidth]{\textit{Trade-off between trust and rate}. The figure corresponds to the hypothetical scenario in which all of the shared keys have the same length (normalised to one) and all $N$ intermediary nodes have the same trust $T$. The lines show the maximum rate achievable such that the total trust $t$ is greater than or equal to a given value. 
    The inset image shows the $N+2$ user network which is fully connected except there is no connection between $A$ and $B$}
    \label{trust}
\end{figure}

The above process results in a final key with length equal to the minimum length of the XORed keys. However it is also possible to consider the scenario in which the end users wish to improve both the security and rate. In this case, after the flooding protocol, the end users partition their set of keys into subsets, indexed by $x$, each comprising of at least three keys, which are subsequently XORed. Each of the resulting keys is then concatenated into a final key. This concatenation further reduces the trust in the final key since all of the combined keys must now individually be secure.  In the non-overlapping case in which there are $m$ subsets, each comprising $n'$ keys such that $n'm=n$, the final trust value is given by:
\begin{equation}
t = \prod_{x}^{m} \bigg(1-\bigg(\prod_{i}^{n'}(1-T_{ix})+\sum_{k}T_{kx}\prod_{i \ i\neq k}^{n'}(1-T_{ix})\bigg)\bigg).
\label{eq:reduced trust}
\end{equation}
We consider as an example the case of a network with $N+2$ users. The network is fully connected except that users $A$ and $B$ do not share a connection. Each of the shared keys between the users are assumed to have the same length (which we normalise to 1) and each of the $N$ intermediary users has the same partial trust $T$. The optimal flooding protocol (either for optimising key rate or security) consists of $N$ non-overlapping paths of the form $A \rightarrow I \rightarrow B$ where $I$ is an intermediary user. In this scenario equation \ref{eq:reduced trust} simplifies to:
\begin{equation}
t=\bigg(\sum_{k=0}^{n'-2}\binom{n'}{k}T^{n'-k}(1-T)^{k}\bigg)^m
\end{equation}
Fig \ref{trust} demonstrates the trade-off between rate and trust for such networks with a variety of network sizes and required trust values.

\subsection{Enhancing security in a realistic quantum network}
\label{sec:FloodingSecurity}
The previous example of implementing the SIAT protocol (as illustrated in Fig. \ref{fig:new user}) assumes a completely trusted third party. In the case of using multiple, partially trusted nodes, flooding can be used (as highlighted previously) to increase the security. It is possible to use flooding to implement the SIAT protocol, in the case that the intermediate nodes are not fully trusted, as we now show.

\begin{figure}[tb]
    \centering
    \includegraphics[width=0.95\columnwidth]{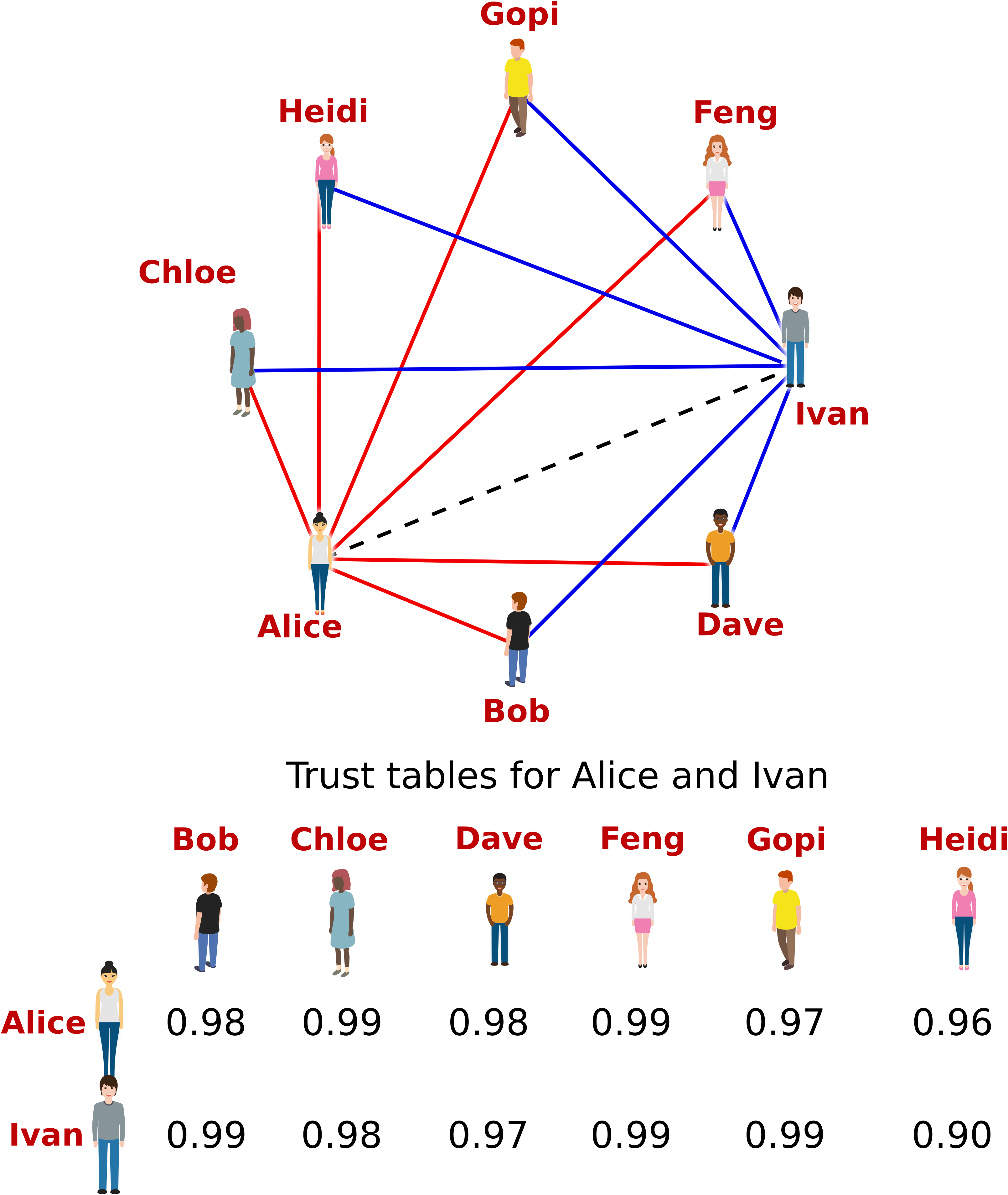}
    \caption{\textit{An example scenario} where Alice (A) does not have a pre-shared authentication key with Ivan (I). Each user has several pre-shared keys with various other users in the network and the amount of trust Alice or Ivan is willing to place in other users is given in the trust table. All links not shown in this example are currently being utilised for other purposes and are thus not available.}
    \label{fig:example}
\end{figure}

Consider the example scenario shown in Fig~\ref{fig:example}. Ivan has been on the network for a while without needing to communicate with Alice, but has been exchanging QKD keys with several other nodes over the network. Ivan publicly announces all (except those he wishes to keep secret for whatever reason) nodes he shares established QKD links with (i.e. a pre-authenticated classical channel and a quantum channel). Alice does the same. In this example they mutually identify 6 nodes they have in common.
Alice and Ivan both have perceptions of how much they personally trust each of these 6 nodes. These partial trust values for both Alice and Ivan are shown in Table \ref{table}.

Through the following steps, we can implement a combined SIAT and flooding protocol (i.e., ``flood the SIAT protocol'') thereby increasing the security of the SIAT protocol.

\begin{enumerate}
    \item Alice and Ivan exchange an inaugural authentication key, via the SIAT protocol using each of the 6 mutual peers. They then XOR these keys to ensure that the inaugural authentication is secure as long as at least two of their mutual peers remained honest (this uses a relaxed definition of `honest' as previously described. We note that security in our protocol would be guaranteed with a minimum of one ideally trustworthy node.). This establishes the dashed link in  Fig~\Ref{fig:example}.
    \item Alice and Ivan share their trust tables (i.e. their partial trust values in each of the other nodes). This can be done publicly or privately using the newly established secure channel.
    \item The minimum of the two trust values assigned by Alice or Ivan for an intermediary node is chosen as the partial trust for that node. 
    \item Alice and Ivan agree upon a minimum end-to-end trust value, say 99.25\%. 
    \item The key flooding protocol is implemented and keys from multiple sets of MNOPs are XORed together until they meet or exceed the minimum end-to-end trust value. If multiple sets of MNOPs can produce keys that exceed the desired trust value then these keys are concatenated together. A simple algorithm (see Section~\ref{sec:floodrouting}) is used to evaluate every possible combination of MNOPs to ensure the best possible final end-to-end key rate given the desired trust threshold.
\end{enumerate}
Using this realistic example we can see that flooding and the SIAT protocol can be easily combined in order to maximise the efficiency and security capabilities of a quantum network. Possible final key rates between end users, and their trust in the security of the key, are shown in Table \ref{table} for the different ways of separating the 6 mutual peers into 2 subnetworks.

\begin{table}[tb]
    \centering
    \includegraphics[width=0.95\columnwidth]{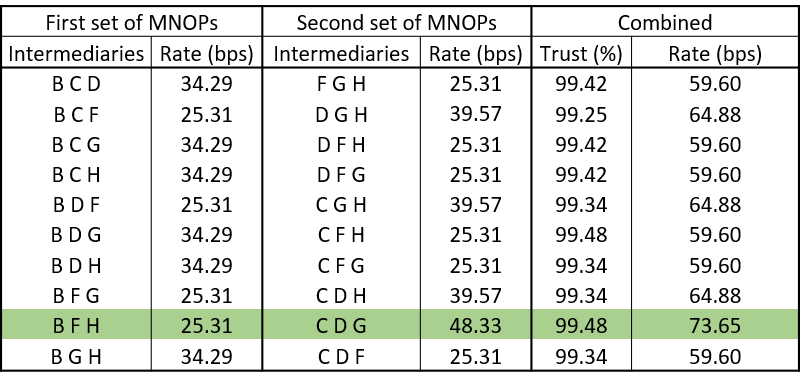}
    \caption{\textit{Demonstrating the use of flooding to improve key rates and security in a realistic network scenario.} Key rate data is taken from our 8-user fully connected quantum test bed and the trust values for the intermediary nodes from Fig \ref{fig:example}. All possible partitions of the 6 intermediary nodes which can simultaneously boost security and rate are shown. Of these, the row highlighted in green provides the best improvement to both security and end-to-end key rate. $A$ and $I$ will also combine their keys obtained via a direct link to further improve the throughput.}
    \label{table}
\end{table}

\section{Conclusion}
Building national or international quantum networks with several users is a labour intensive and costly process. Therefore, it is important to ensure that any network we build can easily be expanded.
Furthermore, the use of trusted nodes in a quantum network is a convenient and effective way to build long distance quantum communication networks. However, the risk that an individual trusted node is compromised remains, so placing absolute trust in any one intermediary node is a serious security flaw. Many quantum networks and communications techniques therefore seek to eliminate trusted nodes. This can be achieved by using measurement device independent QKD \cite{Braunstein2012,Lo2012}; resorting to twin-field techniques \cite{minder2019experimental,lucamarini2018overcoming} (potentially overcoming the secret key capacity for repeaterless QKD protocols \cite{Pirandola2017}); using entanglement distribution~\cite{wengerowsky2018field,wengerowsky2020passively} or by using quantum memories/repeaters~\cite{piparo2017memory,furrer2018repeaters}. However, these solutions are not always applicable.

We have demonstrated a set of algorithmic solutions to growing quantum networks, that, provided the network is large enough and sufficiently well-connected, can be implemented the same way regardless of the size of the network. Following work in Ref. \cite{Salvail10}, we use the concept of partially and temporarily trusted nodes, employing MNOPs to mitigate the associated security risks. 

The SIAT protocol uses trusted nodes to distribute authentication keys to initiate new links, removing the need to physically transport these keys. It shows that, in the case of a well connected network, only $\bigO(n)$ keystores are required, as opposed to $\bigO(n^2)$. It is compatible with a peer to peer like referral scheme, or with centralised authorities. We have experimentally implemented this protocol on our 8-user quantum network test bed and shown that, using just one trusted intermediary node, the SIAT protocol takes between 10 to 40 min to execute. The SIAT protocol can be used in conjunction with classical and/or post quantum authentication protocols like those implemented in Ref~\cite{liu2020experimental}. When using any protocol based on computational security to perform the initial authentication, the probability that such an algorithm can be broken, within the time taken for authentication, upper bounds the value of the partial trust placed in that node.

The SIAT protocol is combined with a flooding protocol, showing how best to calculate which MNOPs to use in any network. We have shown that the larger the network, the larger the gain in security from using MNOPs. The same techniques can be used to improve the end-to-end security and/or the key generation rate between any two end users. In many practical scenarios, as illustrated by the examples we provide, users can choose a good compromise that improves both the security and the final key rate. Furthermore, the flooding protocol can be used to optimise the network performance by using idle network resources to boost network throughput. Additionally by deploying the protocol on our fully connected network, the requirement that an end user know the full network topology and keys lengths is mitigated. So long as there are sufficiently few dishonest users, monogamy of entanglement allows the end users to detect any false reporting of the network topology. 
In a real world implementation users could choose to share information about how much they trust other users openly, via encrypted private channels or using quantum secure anonymous protocols (like Ref.~\cite{huang2020experimental}). Recently it has been suggested that redundant QKD devices, within each individual node, can help overcome certain security concerns~\cite{Zapatero2021}. Our flooding protocol can also be used together with such redundant devices to further optimise both security and key rate.



Together, these two protocols represent a means for quantum networks to be deployed with ease and grow organically according to end user requirements. These protocols are most effective in large densely connected quantum networks where several available MNOPs are likely to exist. The SIAT and flooding protocols presented here can be used in conjunction with most types of QKD protocols (including continuous variable implementations) and hardware as long as MNOPs exist. 
In several cases trusted nodes are the most effective solution and access to these nodes or the amount of memory they have for keystores can be very limited. Quantum communication Cube Satellites  are a good example. 
They are a cheap and effective way to link quantum networks across the globe and almost all such efforts use the satellite as a trusted node~\cite{mazzarella2020quarc,neumann2018q,kerstel2018nanobob}. When building satellite constellations physical access to all optical ground stations to share an initial authentication key is impractical. These protocols would allow authentication transfer between satellites, increase trust levels and help optimally route key generation traffic based on link availability, achieving optimal end-to-end performance for global quantum networks \cite{Harney2021,Chen2021}. Further refinements to these protocols, and secure key storage, would provide complete security solutions for quantum communication networks. 

\clearpage


\section{Methods}\label{sec:methods} 
\footnotesize{
\subsection{Our quantum network test bed}
\label{sec:Net}

We used our entanglement-based quantum network with 8 users~\cite{8user} spread amongst multiple university buildings to implement and test new protocols.
The quantum network architecture is best understood when divided into different layers of abstraction as shown in Fig.~\ref{fig:layers}. The "physical layer" is comprised of hardware necessary to generate, distribute and detect the entangled states, thus forming the actual infrastructure. In this layer, our implementation only requires one single fibre between each user and the (de)multiplexed source, while in the logical/connection layer the topology naturally forms a fully connected graph between all possible pairs formed by the users in the network.
We use one source of polarisation entangled photon pairs and a combination of standard telecom Dense Wavelength Division Multiplexers (DWDM) together with in-fibre beamsplitters (f-BS) in order to distribute bi-partite entangled states between all eight users. 

The multiplexing strategy serves the purpose of fully interconnecting 8 users while only using 16 wavelength channels with 8 f-BS, thus optimising the transmission and entangled state fidelity per channel. Every user is provided with a polarisation analysis module that performs a passive basis choice using a bulk beamsplitter (BS), a half-wave plate (HWP), a polarisation beamsplitter (PBS) and two single-photon  detectors~\cite{8user}.  This enables every user to measure in the horizontal/vertical polarisation basis, or in the diagonal/anti-diagonal polarisation basis when the photons went through the long path with an HWP. 
Our 8 users, referred to as Alice ($A$), Bob ($B$), Chloe ($C$), Dave ($D$), Feng ($F$), Gopi ($G$), Heidi ($H$) and Ivan ($I$), is comprised of two sub-nets of 4 users where each sub-net uses wavelength multiplexing to fully interconnect its members - $A$, $B$, $C$, $D$. The use of f-BS on each of those wavelength channels allows us to duplicate the sub-group creating another set of 4 fully interconnected users - $F$, $G$, $H$, $I$. 

Finally, we require two additional pairs of wavelength to connect the remaining links between the users $AF$, $BG$, $CH$ and $DI$ across the two sub-nets. Any pair of users in the network can perform its own standard BBM92 protocol~\cite{bbm92} where all detected photons from other possible users can be considered as background noise. By choosing a narrow coincidence window (typically $\sim 130 \rm ps$) which can be optimised in post-processing, one can ensure that this noise only increases the Quantum Bit Error Rate (QBER) marginally.
With this setup we can generate secure keys between all 28 possible combinations of paired users. Some extra wavelengths remain, allowing us to form 4 additional links increasing the secret key-rate for some selected users.
\begin{figure}
    \centering
    \includegraphics[width=.9\linewidth]{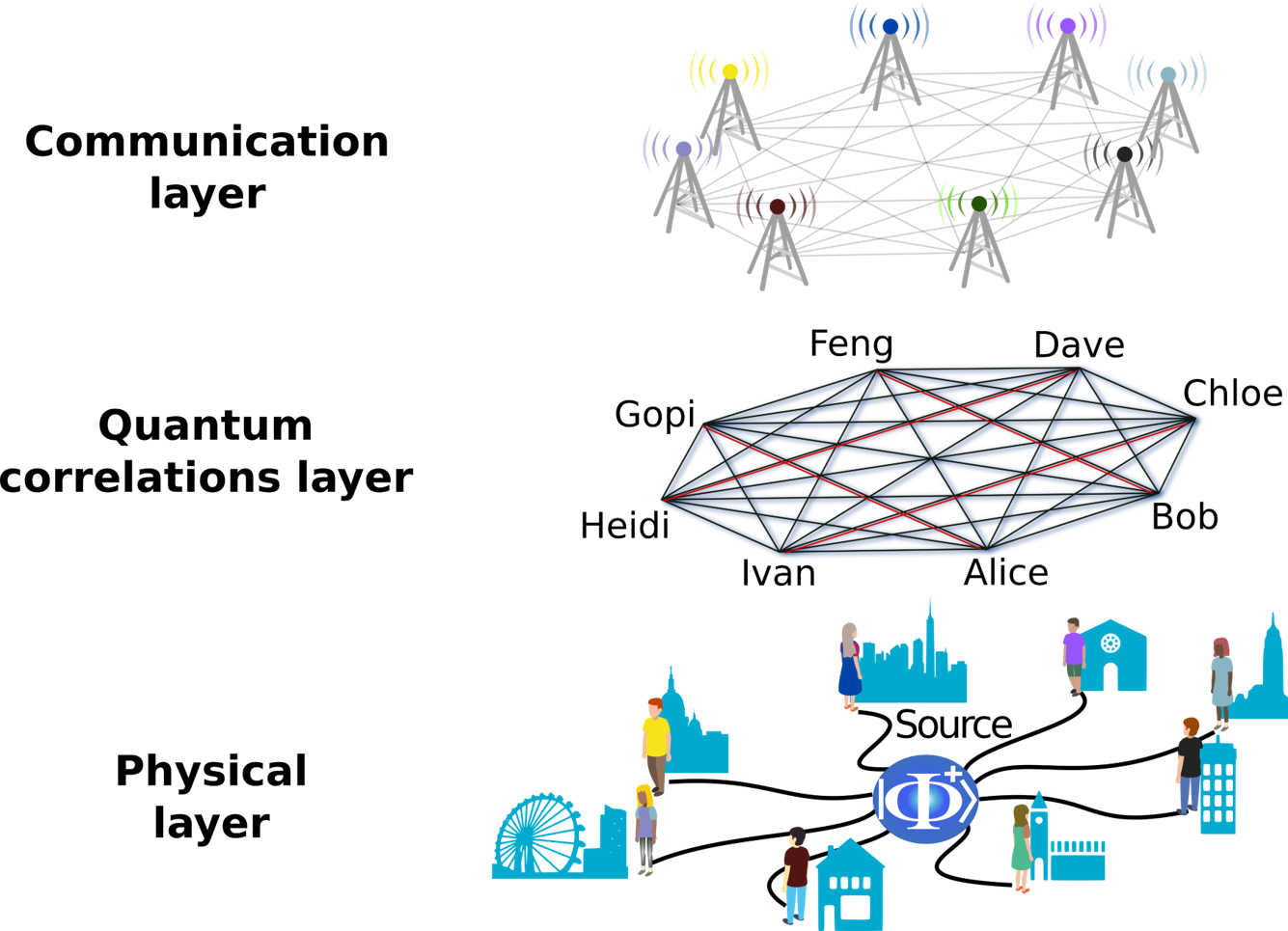}
    \caption{\textit{The network layers}: The Physical Layer represent the physical hardware including the photon pair source, the optical components necessary to multiplex and distribute the photons toward the users together with their detection analysis modules. 
    The quantum correlation layer represents the sharing of entanglement between users  and the communication layer is where the measurement results are processed to implement QKD, Flooding and/or the SIAT protocols.
    }
    \label{fig:layers}
\end{figure}
\subsection{The network service provider}
We use data from the experiment described in \cite{8user}. Our network architecture relies on a Quantum Network Service Provider (QNSP) to distribute bipartite polarisation entangled states to users with Polarisation Analysis Modules (PAM). The QNSP is comprised of an entangled photon pair source employing a Sagnac scheme and a Multiplexing Unit (MU). The Sagnac source consist of a 5 cm-long Magnesium Oxide doped periodically poled Lithium Niobate (MgO:ppLN) bulk crystal with a poling period of 19.2\,$\mu$m which is pumped by a CW laser emitting at $775.1085 \rm{~nm}$ in both directions inside the Sagnac loop. The input/output of this loop is defined by a dichroic mirror and a polarisation beamsplitter (PBS) allowing for diagonally polarised pump light to split and propagate the horizontally (vertically) polarised part anti-clockwise (clockwise) inside. An half-wave plate (HWP) after the PBS transmission port is used to set the pump light in the anti-clockwise to vertical and also allows for the $|V_s\rangle|V_i\rangle$ signal and idler photons pairs generated through type-0 spontaneous parametric down-conversion in the other direction to become $|H_s\rangle|H_i\rangle$. This make it possible for both contributions in the loop to coherently recombine at the PBS and exit isolated from pump light by the dichroic mirror. 
The $|\Phi^+\rangle$ Bell state generated spans 32 channels in the C-band.
The distribution of these 32 channels (as defined by the International Telecommunication Union in G.694.1)  in each of the fibres was achieved by our MU consisting of a set of 8 standard 50:50 fused couplers with insertion loss below $3.4 \rm{~dB}$, together with 16 add/drop thin-film DWDMs showing ~$0.5 \rm{~dB}$ insertion loss and a channel spacing and nominal full width of $100 \rm{~GHz}$. Fibre Polarisation Controllers (FPCs) were used to ensure that the reference frame of polarisation in the source is (nearly) identical to that of the PAM.
\subsection{Data generation \& analysis}
The data was generated during the operation of the quantum network over a span of several days. The duration of each data collection run was limited for logistical reasons and/or the hold time of the cryostat used for the detectors ($\approx$18 hours). At the beginning of each data collection run, we polarisation neutralised all fibres. For fibres that carry several wavelengths, they were neutralised at their central wavelength. We also tuned the state produced by the source to minimise the QBER for one of the connections. Signals from all 16 detectors were collected by a Swabian Instruments time tagger and these data were stored. While processing the data, each user's counts were extracted from the data file and stored into 8 separate files. These were then processed to generate the final key rates. The rates used here account for finite key effects.

We processed the data in 20 minute blocks, the first few seconds of data of each block were used to compute the optimal coincidence window to use for that block. 

We used the Wegman-Carter authentication scheme~\cite{Wegman1981NewEquality} to estimate the amounts of authentication key needed and the average key rates generated by our real network over 18 hours to simulate the running of these protocols.

\subsection{Calculating the optimal flooding protocol(s)} \label{sec:floodrouting}
The network can be represented by a simple graph where two vertices $\boldsymbol{v_{i}}$ and $\boldsymbol{v_{j}}$ are connected by an edge $(\boldsymbol{v_{i}},\boldsymbol{v_{j}}) \in E$ if they share a key $k_{ij}$. Each edge is assigned a capacity which corresponds to the length of the key shared between the users per unit time (i.e. the key rate). In the case of our quantum network test bed this corresponds to the length of key generated between two users in each 20 minute block. This gives rise to a symmetric capacity matrix $\mathcal{C}$ where $\mathcal{C}_{ij}$ is the capacity of the edge connecting the vertices $i$ and $j$.

Given two parties who wish to generate a key across the network, we consider one, Alice, acting as a source and the other, Bob, acting as a sink. As discussed in Sec \ref{key flooding} in terms of the graph view, a given flooding protocol $p_{i}$ corresponds to an assignment of an orientation to all of the intermediary edges. This orientation imposes a partial time ordering on the vertices in which a vertex $i$ acts before another $j$ if there is a positively orientated edge ($i\rightarrow j)$ connecting them. Orientating the edges transforms the simple graph $\mathcal{N}$ into a directed graph $\mathcal{N}_{p_{i}}$. Each possible directed graph corresponds to a set of flooding protocols which are equivalent up to the ordering of the output edges when the key splitting is undertaken (as discussed in Sec \ref{sec:flood_protocol}). 

Each flooding protocol has a maximum length of secure key that can be generated between Alice and Bob. This length corresponds to the maximum flow between the two vertices representing the parties in the directed graph. Since each directed graph in general has its own maximum flow we refer to the directed graph maximising the maximum flow as the optimum directed graph and the corresponding maximum flow as the optimum flow. A flooding protocol achieving the optimum flow is called the optimum flooding protocol(s).

The maximum flow between two vertices in a directed graph is well known to be related to minimum cuts in the network by the max-flow min-cut theorem \cite{Ford1956}. A cut is a bi-partition of the vertices of the graph such that the source and sink lie in different sets. The cut-set $\Tilde{C}$ consists of all edges passing across the cut. Under a multi-path routing strategy such as flooding the maximum flow $F_{max}$ is given by:

\begin{equation}
F_{max}(\mathcal{N}_{p_{i}})= \underset{\Tilde{C}}{min} \sum_{(i,j) \ \in \ \Tilde{C}} \mathcal{C}_{ij}
\end{equation}
which is the sum of the capacities of all the edges passing through the cut with the minimum value. 

This can be implemented computationally using the Edmonds-Karp algorithm \cite{Edmonds1972} to find the optimum flooding protocol(s). We first convert the undirected graph representing our network into a directed multi-graph in which each edge in the original graph is converted into two edges with opposite orientations. We then perform a breadth-first search of the graph beginning from the source vertex and ending with the sink. By back-tracing it is possible to find an augmenting path from the source to the sink with available capacity. We then send the maximum possible flow along this path and remove this from the capacity matrix to find the residual graph. This process continues until there are no further paths with available capacity and the flow is maximised. An optimum flooding protocol (and corresponding optimum directed graph) can be found from the orientation of edges in the augmenting paths. More efficient algorithms such as Orlin's algorithm \cite{Orlin2013} can reduce the run time to $\mathcal{O}(VE)$ where $V$ and $E$ are the number of vertices and edges respectively.

The case in which we consider partial trust is slightly more complex as it may be possible that no optimum flooding protocol satisfies the trust requirements of the end users. However a different flooding protocol, corresponding to a different directed graph, may satisfy the trust requirements. In this case it is necessary to pre-define the orientation of the edges, converting the graph into a directed graph $\mathcal{N}_{p_i}$. The Edmond's Karp algorithm may then be used to find the maximum flow for this flooding protocol. The optimum protocol which obeys the end-users trust requirements may then be found by performing the procedure for all possible directed graphs and calculating the end-to-end trust values. We note that it may be more efficient to first calculate which combinations of MNOPs satisfy the trust requirements and then restrict to calculating the maximum flow for directed graphs which contain these paths. 

Finally we note that an optimum flooding protocol is in general non-unique. This non-uniqueness may arise in two ways: at the directed graph level and at a 'key splitting' level. Different directed graphs with the same capacity matrix may achieve the same optimum flow in which case there is more than one time ordering of the intermediary nodes that can perform an optimal flooding protocol. Additionally even for the same directed graph and ordering of intermediaries there may be more than one optimal protocol. This arises due to the different ways keys  may be split as discussed in Sec \ref{sec:flood_protocol}.

\subsection*{Data availability}
All data from this publication is stored for at least 10 years on the University of Bristol's Research Data Storage Facility (RDSF). The processed data for the findings in this paper is available publicly from the RDSF, the raw data consisting of timetag files is too large to host publicly and available from the corresponding author on request.
}
\subsection*{Acknowledgements}
The research leading to this work has received funding from United Kingdom Research and Innovation's (UKRI) Engineering and Physical Science Research Council (EPSRC) Quantum Communications Hub (Grant Nos. EP/M013472/1, EP/T001011/1) and equipment procured by the QuPIC project (EP/N015126/1). We also acknowledge the Ministry of Science and Education (MSE) of Croatia, contract No. KK.01.1.1.01.0001. We acknowledge financial support from the Austrian Research Promotion Agency (FFG) project ASAP12-85 and project SatNetQ 854022. SP acknowledges support from the European Union via \textquotedblleft Continuous Variable Quantum Communications\textquotedblright\ (CiViQ, Grant agreement No. 820466). NRS was funded by the EPSRC through the Quantum Engineering Centre for Doctoral Training, EP/SO23607/1. AF was funded by the EPSRC via a Doctoral Training Partnership EP/R513386/1. We would like to thank Thomas Scheidl for help with the software used to run the original experiment and Mohsen Razavi \& Guillermo Curr\'as Lorenzo for their help proving the security of the implementation of the original network experiment. We would also like to thank Rui Wang for noticing and helping correct an error in the original version of the paper.


\subsection*{Author Contributions}
NRS and SKJ conceived the authentication protocol, AF and SP implemented the flooding protocol. NRS, AF, SKJ, DA and SP explored the relationships between these two protocols and the use cases. NV and BL wrote the post-processing software. SKJ, DA, SW, ML,SPN, BL, ZS, MS, and RU performed the quantum network experiment. SKJ, SP and JGR supervised the project and contributed to the ideas. SKJ was the team leader. The paper was written by NRS, AF, DA and SKJ and all authors read and contributed to improving it.

\bibliography{Qnet}
\clearpage
\appendix


\section*{Supplementary Information}\label{appendix}
\renewcommand{\figurename}{Supplementary Figure}
\renewcommand{\tablename}{Supplementary Table}

\subsection{Derivation of authentication key scaling - Eq. \ref{eq: scaling}} \label{sec: scaling deriv}

Equation \ref{eq: scaling} shows the minimum numbers of keys that must be shared while initiating a fully connected, authenticated network, in which there is the possibility of up to $c_a$ users forming a collective adversary:$$ n_k = \frac{(c_a+2)(c_a+1)}{2} + (n-c_a-2)(c_a+2).$$

As discussed, to ensure the security of the SIAT protocol when sharing authentication keys, there must be $c_a + 1$ disjoint paths between every pair of users that with to communicate (or a direct path). We assume that every pair of users wish to communicate.

For $n>c_a + 1$, consider adding a new user. They must have received at least $c_a + 1$ pre-shared keys in order to have authenticated communication with every other user; in fact, they need to receive that many exactly, if the underlying network is fully connected - that is, for each user with whom they share a key, they know that user has an authenticated channel with any other user with whom they may wish to communicate.

When considering the total number of pre-shared keys required, we first need to construct a fully connected network for the initial $n = c_a + 1$ users, which requires $\frac{1}{2}(c_a + 2)(c_a + 1)$ pre-shared keys. Every remaining user needs to share keys with $c_a + 1$ of these users, which means an additional $(n - c_a - 1)(c_a + 1)$ keys are required.


A similar formula can be derived on need for any desired network topology.

\subsection{Calculating trust duration when adding users into the network - Fig. \ref{fig:new user}}\label{sec:adding user}

The WC authentication protocol \cite{Wegman1981NewEquality}, which is the most widely used, makes use of hashing functions which give each message a tag dependent on the message and a pre-shared key.

In the scheme, Alice sends Bob a message $m$ out of a set $M$ of possible messages, and appends the tag $f(m)$, in which $f$ is a function from the set $F$ which maps the set $M$ to the set $T$ of possible tags. In order for Bob to verify that this tag is genuine and the message was sent by Alice, Alice and Bob must have an initial shared key ($k_{\text{Auth}}$) which specifies which member of $F$ is used for the tag. Therefore, the length of $k_{\text{Auth}}$ depends on the size of $F$ required.

Say that we would like the scheme to be insecure by probability $c$. Ref. \cite{Wegman1981NewEquality} shows that the authentication tag must be in a set of size $|F|$, where $|F| = \frac{2}{c}$. Hence the tag length, $b$, should be $\log_2 (\frac{2}{c})$ bits. 

Ref. \cite{Wegman1981NewEquality} then shows that the length of the original shared key should have length $s = 4(\log_2 (\frac{2}{c}) + \log_2 \log_2(d))\log_2(d)$ in which $d$ is the length, in bits, of the message $m$. We assume that this is proportional to the final key length, $a$, i.e. $ga$ for some constant $g$. Experimentally, this is the amount of classical authenticated data that the two users need to exchange to be able to generate each bit of key. We are then interested in the number of bits of authentication key required per bit of final secret key, for insecurity $c$ and final key length $a$, i.e. the proportion $$4(\log_2 (\frac{2}{c}) + \log_2 \log_2(ga))\log_2(ga)/a.$$ This is the proportion of key produced in each round of QKD that should be used as the authentication key for the following round. This is decided either arbitrarily, or dependent on the length of each round, $a$; conservatively let's say 10\% of each round is reused for authentication.

We then solve $0.1 = 4(\log_2 (\frac{2}{c}) + \log_2 \log_2(ga))\log_2(ga)/a$ for $a$ to find the ideal number of bits to be produced in each round. For every event detected by a user, the arrival time is measured by the time tagger and stored in a 64 bit binary format. Further the basis information (1 bit for each detected photon) is also recorded. Lastly, information is exchanged back and forth to perform error correction, status checks, etc. In our experiment this was at most 7 more bits of classical information sent between the users per photon detected. Given the experimentally measured key rates and detection rates, we conservatively estimate that for every every bit of key generated we had at most 10000 individual photon detection events. Thus we chose a value of $g$ of 720 000. In this case, we find each round should be 50 563 bits, and so the key length should be $s = 4(\log_2 (\frac{2}{c}) + \log_2 \log_2(d))\log_2(d) =2 179$ for our desired valued of $c$, $10^{-9}$. 

Fig. \ref{fig:new user} shows the amount of time needed to carry out the SIAT protocol - that is, the amount of time needed (with previously experimentally generated key rates) for the trusted node to send the new user and the desired end user the 5056-bit tag, and then the amount of time needed for the new user and the end user to carry out a full 50 563-bit round of QKD.

For example, let's say the user I would like to make a connection with the user $A$, and already has a connection with the trusted node $B$. The mean key rate for connection $AB$ in \cite{8user} is 45.94 bits per second, and 34.70 bps for $BI$. For the first part of the SIAT protocol, the trusted node $B$ sends a 5056-bit tag to $A$ and $I$ simultaneously, which takes $5056/34.70 = 145.71$ s. $A$ and $I$ then carry out a full 50563-bit round of communication. As the mean key rate in communication between $A$ and $I$ is 28.52 bps, this takes 1773\,s, and the total required time is 1918\,s ($\approx$ 32 minutes). This is represented by the first bar in Fig. \ref{fig:new user}.

We note that a similar process can be used to add users into any quantum network where the topology allows the end users to share initial authentication keys with at least one common node or set of nodes.

\subsection{Announcements for optimal flooding with fully trusted nodes}
\label{appendix:comic}
\begin{figure*}[tb]
    \centering
    \adjustbox{trim=0cm 0cm 0cm 0cm, clip}{\includegraphics[width=0.95\linewidth]{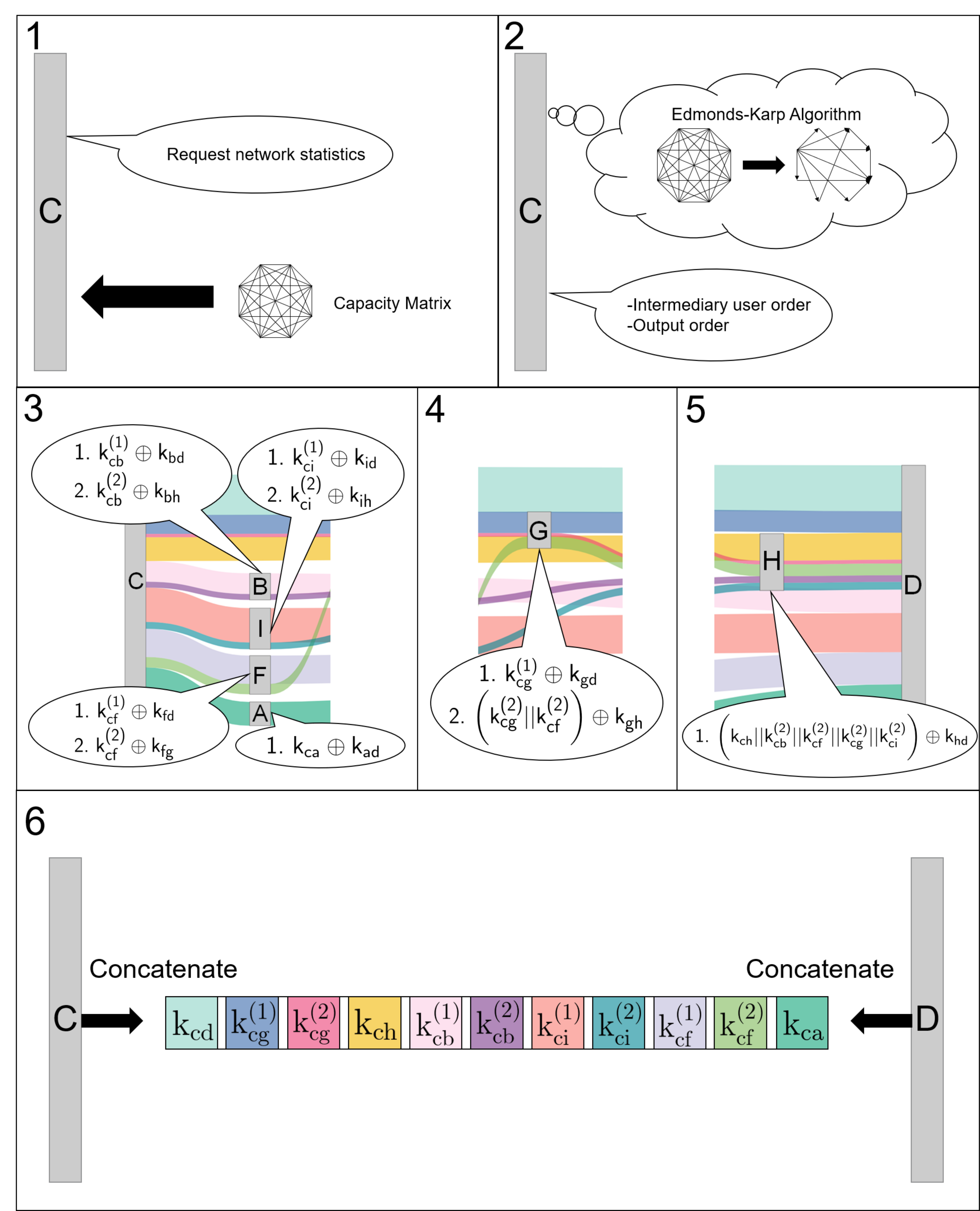}}
    \caption{\textit{Step by step implementation of our flooding protocol on the quantum network test bed}: Panel 1 shows user $C$ requesting network statistics to determine the capacity matrix of the available network resources. Panel 2 shows user $C$ finding an optimal quantum secure flooding protocol using the Edmonds-Karp Algorithm and the capacity matrix. User $C$ communicates to the intermediary nodes their ordering and the order in which to use their remaining keys to output their received keys. Panels 3, 4 and 5 show the announcements made by the intermediary nodes. Panel 6 shows the two end users privately concatenating their shared keys.}
    \label{comic}
\end{figure*}

Figure \ref{sankey trusted} shows the optimum end-to-end flooding protocol between users $C$ and $D$. The announcements made by the intermediary nodes were suppressed in the figure to preserve clarity. Figure \ref{comic} shows the full set of announcements made by the intermediary nodes to pass knowledge of keys from $C$ to $D$. These users concatenate these keys to form the final flooded key $k_{cd}^{\mathrm{(flood)}}$ as shown in Fig. \ref{sankey trusted}. 

\clearpage

\end{document}